# Physics-Informed Residual Deep Learning for Constitutive Modeling of Hot Deformation and Dynamic Recrystallization in a Mo-Rich α+β Titanium Alloy

Prashil S. Joshi[1], Diksha Mahadule[2], Rajesh K.Khatirkar[2*]

(1) Prashil S. Joshi[1]

E-mail: prashiljoshi@iisc.ac.in

Affiliation(s): [1]Department of Materials Engineering, Indian Institute of Science, Bengaluru, Karnataka - 560012, India.

ORCID: orcid.org/0009-0009-6727-5483

(2) Diksha Mahadule[2]

E-mail: diksha.mahadule7@gmail.com

Affiliation(s): [2]Department of Metallurgical and Materials Engineering, Visvesvaraya National Institute of Technology (VNIT), South Ambazari Road, Nagpur – 440010, Maharashtra, India.

ORCID: orcid.org/0000-0002-7596-5830

(3) Rajesh K. Khatirkar[2] (***Corresponding author**)

E-mail: rajesh.khatirkar@gmail.com, rajeshk@mme.vnit.ac.in

Affiliation(s): [2]Department of Metallurgical and Materials Engineering, Visvesvaraya National Institute of Technology (VNIT), South Ambazari Road, Nagpur – 440010, Maharashtra, India.

Phone: +91-712-2801508, Fax: +91-712-2223230

ORCID: orcid.org/0000-0001-6421-9368

**Abstract**

Accurate constitutive modeling of hot deformation characteristics is essential for the development of thermomechanical procedures for advanced structural alloys. Conventional Arrhenius-type and empirical models frequently find it challenging to effectively depict the combined influences of strain hardening, dynamic recovery (DRV), and dynamic recrystallization (DRX) across extensive processing intervals. This research uses two Stacked Residual Physics-Informed Neural Networks (STAR-PINNs) to predict the hot-deformation characteristics of a Mo-rich α+β titanium alloy (Ti–6Al–4Mo–1V–0.1Si). The Improved STAR-PINN integrates thermomechanical constitutive restrictions, while the DRX-Aware STAR-PINN utilizes a dual-output framework that incorporates recrystallization kinetics through a trainable JMAK–Avrami model. Both models utilize a common residual encoder that is trained on empirical flow-stress data collected at temperatures ranging from 800 to 1050 °C and strain rates between 0.01 and 10 $s^{-1}$. Constraints informed by physics that enforce thermal softening, strain-rate sensitivity, strain hardening, and post-peak softening are applied via automatic differentiation. The DRX-Aware model also includes JMAK–Avrami regularization featuring learnable, constrained kinetic parameters k and n, DRX saturation limits, Arrhenius-based consistency, and latent-feature integration connecting the estimated DRX fraction to the stress output. The DRX-Aware STAR-PINN obtained RMSE = 11.62 MPa, MAE = 4.73 MPa, and $R^2$ = 0.9852 on the test set reserved for validation, along with a five-fold cross-validated RMSE of 12.74 ± 0.38 MPa. The learned JMAK parameters converged to $k = 2.03$ and $n = 1.99$, remaining close to the conventional values while being inferred directly from the data. The model effectively mirrored flow curves influenced by temperature, DRX development, and Zener–Hollomon correlations under interpolation scenarios while preserving physically coherent constitutive behavior. These findings illustrate that physics-informed deep learning delivers a precise and understandable structure for constitutive modeling, presenting a viable approach for sophisticated process modeling of hot-deformed titanium alloys.

## 1. Introduction

Titanium alloys occupy a position of critical importance in advanced engineering applications owing to their exceptional combination of high specific strength, outstanding corrosion resistance, and biocompatibility [1–3]. The aerospace sector relies extensively on titanium components for airframe structures, compressor blades, and fasteners, where weight savings directly translate to fuel efficiency gains and payload improvements [4,5]. The ongoing drive for elevated operating temperatures and intricate structural designs has increased the need for $\alpha+\beta$ and near-$\beta$ titanium alloys with regulated microstructures. Mo-containing titanium alloys, specifically, have garnered significant research attention due to the potent $\beta$-stabilizing influence of molybdenum, which elevates the $\beta$-transus temperature, facilitates retention of the $\beta$-phase, and allows for wider hot-working ranges compared to traditional Ti–6Al–4V [4,6–8]. Realizing the full potential of these alloys requires a detailed understanding and an accurate mathematical description of their hot-deformation response across the entire thermomechanical processing space.

Hot deformation of titanium alloys involves the simultaneous operation of competing microstructural mechanisms, including dislocation generation and interaction (work hardening), thermally activated dislocation climb and annihilation (dynamic recovery, DRV), and nucleation and growth of strain-free grains (dynamic recrystallization, DRX) [9–12]. The relative dominance of these mechanisms is governed by the Zener–Holloman temperature-compensated strain-rate parameter $Z = \dot{\varepsilon}\cdot\exp(Q/RT)$, $\dot{\varepsilon}$ = strain rate ($s^{-1}$), Q = activation energy for deformation (J/mol), R = universal gas constant (8.314 J/mol·K), T = absolute temperature (K), which collapses the influence of temperature and strain rate into a single thermally activated kinetics descriptor [10,13]. At high temperatures and low strain rates (low Z), DRX predominates and produces characteristic flow softening after a peak stress; at low temperatures and high strain rates (high Z), work hardening and DRV compete without significant recrystallization [14]. Accurate prediction of flow stress across this mechanistic spectrum is essential for finite-element simulation of forging, rolling, and extrusion operations. Classical constitutive frameworks for hot deformation fall into three broad categories. The empirical Johnson–Cook (JC) model [15] describes the flow stress as a multiplicative function of plastic strain, strain rate, and temperature, but its decoupled form fails to capture the complex interactions among these variables during DRX. The phenomenological Arrhenius sinh-type equation introduced by [11,16] and subsequently refined by [9] relates the Zener–Hollomon parameter to flow stress via hyperbolic-sine kinetics and has been widely applied to titanium alloys [17,18]. While the Arrhenius approach has a rigorous physical basis rooted in thermally activated deformation theory [13] it is fundamentally a single-valued steady-state description that is incapable of reproducing the transient peak–softening–steady-state flow curve morphology characteristic of DRX [19,20]. Modified Arrhenius equations with strain-dependent material constants have been proposed [21] but the polynomial fitting required introduces overfitting risks and provides no mechanistic connection to the underlying microstructural evolution.

The limitations of classical constitutive models have motivated the application of machine learning methods, particularly artificial neural networks (ANNs), to constitutive modeling [22]. [23] pioneered the use of neural networks for material behavior characterization, and subsequent work demonstrated their capacity to approximate highly nonlinear constitutive relationships without explicit functional assumptions [24,25]. For titanium alloys specifically, ANN models have been shown to outperform Arrhenius equations in reproducing flow curves across wide thermomechanical windows [26,27] demonstrated that ANN constitutive models achieve higher accuracy than Arrhenius-type models for high-temperature deformation of Aermet100 steel, and analogous results have been reported for magnesium [28] and near-β titanium alloys [29]. However, conventional ANN models lack physical interpretability: they function as black-box function approximators that may produce predictions violating fundamental metallurgical principles when evaluated outside the training domain [30,31]. This physical inconsistency is particularly problematic for industrial applications, where constitutive equations must remain reliable under extrapolation to process conditions not represented in the training data.

Recent developments in high-performance computing, reduced-order modeling, and data-driven materials science have facilitated the creation of extensive constitutive datasets appropriate for machine learning uses. Deep learning methods have shown exceptional ability to accurately and efficiently capture complex history-dependent plastic deformation behavior, providing a viable alternative to traditional constitutive modeling frameworks [32,33]. Even with these improvements, solely data-driven models frequently lack physical interpretability and might breach established metallurgical principles beyond the training domain. Physics-informed neural networks (PINNs), introduced by [34], offer a principled solution by incorporating governing physical equations directly into the neural network loss function through automatic differentiation. By penalizing violations of physical constraints as additional loss terms, PINNs are encouraged to satisfy known governing equations simultaneously with fitting the training data, thereby achieving both data-driven accuracy and physical consistency [35,36]. The PINN framework has been successfully applied to solid mechanics inverse problems [37], high-speed fluid dynamics [38], material characterization from indentation [39], and microstructural defect analysis [39]. Crucially, PINNs trained with physical penalty are, in principle, expected to retain their physical constraints under extrapolation, and are therefore expected to generalize more reliably than purely data-driven models for out-of-distribution prediction [40,41]. This potential advantage, however, requires explicit evaluation under conditions withheld entirely from training, since satisfying a physics-informed penalty on the training data does not by itself guarantee accurate behaviour outside the sampled processing window.

Despite the clear relevance of this framework to constitutive modeling, the systematic incorporation of hot deformation physics, including hardening, thermal softening, strain-rate sensitivity, and DRX, into PINN loss functions for titanium alloys remains underexplored. The deep learning architecture underpinning the present models is a Stacked Residual Neural Network (STAR-NN) built upon the residual learning principle introduced by [42,43]. Residual connections provide direct gradient pathways through the network, alleviating the vanishing-gradient problem that limits the depth and expressive capacity of conventional feedforward architectures [44,45], Layer normalization [46] and SiLU (Sigmoid Linear Unit) activation functions [47] are employed within each residual block, where SiLU's smooth differentiability at all points is particularly important for physics-informed training, which requires second-order automatic differentiation of the network output to compute curvature penalties. The AdamW optimizer [48] with decoupled weight decay and cosine annealing learning-rate scheduling [49] were employed for all models.

Predictive uncertainty quantification is an increasingly important requirement for engineering constitutive models, particularly when used to drive process optimization under uncertainty [50,51]. Monte Carlo Dropout (MC-Dropout), introduced by Gal and Ghahramani [51] approximates Bayesian inference by performing multiple stochastic forward passes through a network with dropout active at inference time. The resulting distribution of predictions provides per-input uncertainty estimates without the computational overhead of full Bayesian training [50,52]. A well-calibrated uncertainty estimate is critical for the downstream use of constitutive models in process control applications, where regions of high model uncertainty correspond to processing conditions that require additional experimental validation.

DRX modelling has traditionally relied on phenomenological JMAK (Johnson–Mehl–Avrami–Kolmogorov) kinetics [53,54] which describe the evolution of recrystallized volume fraction as a sigmoidal function of normalized strain [55,56]. The JMAK equation is physically grounded in nucleation and growth statistics but requires fitting of kinetic parameters that vary with deformation conditions[16,57]. Coupling JMAK kinetics with constitutive predictions has been achieved through multi-equation systems in the classical modeling literature [58,59], but this approach requires sequential solution of coupled equations and assumes the JMAK functional form is valid across the entire processing space. The DRX-Aware STAR-PINN proposed in this work circumvents these limitations by jointly training stress and DRX fraction predictions within a single neural network framework, using JMAK kinetics only as a soft regularization prior while retaining data-driven flexibility to capture deviations from ideal kinetics [60]. Multi-task learning through parameter sharing has been shown to improve generalization by regularizing shared representations through diverse but related objectives [60], providing a theoretical motivation for the joint prediction architecture.

[61] recently presented the STAR-PINN architecture to tackle nonlinear magnetic diffusion issues, showing that residual-stacked physics-informed networks can successfully capture intricate nonlinear material behaviors while ensuring physical consistency. Inspired by its capacity to depict highly nonlinear constitutive behavior, the STAR-PINN framework was enhanced in this study for physics-informed modeling of hot deformation and dynamic recrystallization processes in a Mo-rich α+β titanium alloy [6–8,22]. First, a residual PINN architecture (Improved STAR-PINN) is developed that enforces five thermomechanical constitutive constraints which are thermal softening, strain-rate sensitivity, pre-peak hardening, post-peak softening, and curvature regularization, simultaneously via automatic differentiation and is validated on Ti–6Al–4Mo–1V–0.1Si across a wide thermomechanical window. Second, a novel DRX-Aware STAR-PINN is introduced in which the DRX fraction output is explicitly concatenated with latent features before stress decoding, encoding the mechanistic pathway recrystallization → flow softening directly into the network topology. Third, a DRX–softening coupling loss is proposed that penalizes simultaneous positive gradients of both the DRX fraction and stress with respect to strain, providing gradient-level enforcement of the anti-correlation between the microstructural state and macroscopic flow resistance. Fourth, both models are systematically evaluated using five-fold cross-validation, MC-Dropout uncertainty quantification, and comprehensive error analysis across the full processing space.

## 2. Experimental Procedure and STAR-PINN Framework Description

### 2.1 Experimental Procedure

A comprehensive hot deformation investigation was carried out on a Mo-rich α+β titanium alloy, Ti–6Al–4Mo–1V–0.1Si (wt.%), to establish a robust experimental database for the development of physics-informed machine learning-based constitutive models. The alloy was synthesized through vacuum induction melting followed by sequential forging and hot rolling operations to obtain sheets with a final thickness of approximately 8 mm. The chemical composition of the alloy was measured using optical emission spectroscopy (OES) and was determined to be Ti–88.1%, Al–6.18%, Mo–3.96%, V–1.36%, Si–0.10%, and O–0.0164% (wt.%). Based on metallographic observations and empirical composition-based relations available in the literature, the β-transus temperature of the alloy was estimated to be approximately 965°C. Prior to deformation testing, the as-rolled sheets were homogenized at 950°C for 45 min. The homogenization cycle was repeated three times with intermediate furnace cooling in order to minimize residual chemical segregation and microstructural heterogeneity arising from prior thermomechanical processing. This treatment produced a relatively uniform duplex (α+β) microstructure suitable for high-temperature deformation studies. Cylindrical hot compression specimens with dimensions of 6.0±0.05 mm diameter and 9.0±0.05 mm height were machined using electrical discharge machining (EDM), with the specimen axis oriented parallel to the rolling direction. The specimen end faces were carefully polished to ensure parallelism and minimize barreling effects during deformation. Hot compression experiments were conducted using a DARTEC 100 kN servo-hydraulic testing system equipped with a high-temperature resistance furnace. Before deformation, each specimen was soaked at the target temperature for 15 min to ensure thermal equilibrium throughout the sample volume. Compression testing was performed up to a true strain of approximately 0.7, corresponding to nearly 50% height reduction. The deformation temperatures ranged from 800 to 1050 °C at intervals of 50°C, while the strain rates varied between 0.01 and 10 $s^{-1}$ [6–8]. Consequently, 30 independent thermomechanical deformation conditions were investigated. Graphite lubricant was applied at the die–specimen interface to reduce friction and promote homogeneous deformation. The load–displacement responses acquired during testing were converted into true stress–true strain data using standard constitutive relationships. After deformation, all specimens were water-quenched to preserve the deformation-induced microstructures. Metallographic characterization was subsequently carried out on sections parallel to the compression axis. Surface preparation involved sequential grinding with SiC papers, followed by polishing with alumina and diamond suspensions to achieve a mirror-like finish. The microstructures were revealed using Kroll's reagent and analyzed using a JEOL JSM-6380A scanning electron microscope operating in secondary electron mode. The experimentally obtained stress–strain responses exhibited characteristic features associated with hot deformation of titanium alloys, including strain hardening, dynamic recovery, flow softening, and dynamic recrystallization. These experimentally measured constitutive responses formed the basis for the development and validation of the proposed STAR-PINN frameworks.

## 2.2 Dataset Preparation for Physics-Informed Constitutive Learning

Hot-compression experiments were conducted on a Mo-rich α+β titanium alloy (Ti–6Al–4Mo–1V–0.1Si) over 800–1050 °C and 0.01–10 $s^{-1}$ strain rates, yielding ~7,100 experimental observations. Each data point is described by three inputs and one output:

$$x = [\varepsilon, \ln(\dot{\varepsilon}), T] \rightarrow \quad y = \sigma \text{ (MPa)}$$

All inputs were independently normalized to [0, 1] using min–max scaling, with the scaling parameters calculated from the training set only. The dataset was divided into 80% training (5694 observations) and 20% hold-out test (1424 observations). Model robustness was additionally assessed using five-fold cross-validation. For this analysis, the five folds were generated from the complete dataset, including observations from the hold-out partition; therefore, the five-fold cross-validation and the 80/20 hold-out evaluation represent separate, non-nested assessments and should not be interpreted as a strictly independent validation of the same test set. Peak stress $\sigma_p$ = 420.33 MPa occurs at peak strain $\varepsilon_p$= 0.1459. The DRX critical strain was estimated as $\varepsilon_c = 0.6\varepsilon_p = 0.0876$, consistent with the Poliak–Jonas criterion.

### 2.2.1 Motivation for Physics-Informed Learning

Unlike conventional neural networks that rely solely on data fitting, the proposed framework embeds metallurgical knowledge directly into the optimization process. Physics constraints were imposed through automatic differentiation. The following physically meaningful trends were enforced:

$$\frac{\partial \sigma}{\partial T} < 0 \text{ (thermal softening),}$$

$$\frac{\partial \sigma}{\partial \ln(\dot{\varepsilon})} > 0 \text{ (strain-rate sensitivity),}$$

$\frac{\partial\sigma}{\partial\varepsilon} > 0$ before DRX initiation (strain hardening), and

$\frac{\partial\sigma}{\partial\varepsilon} < 0$ after the critical strain $\varepsilon_c$ (flow softening).

These constraints guide the model toward thermodynamically consistent constitutive behavior.

| Constraints | Physical basis |
|---|---|
| Thermal softening | Thermal activation reduces dislocation resistance |
| Strain-rate sensitivity | Less recovery time at higher strain rates |
| Strain hardening | Dislocation accumulation before DRX onset |
| DRX softening | Recrystallized grains reduce dislocation density |

*Table 1: Physical basis for the constraints imposed in the model.*

These constraints were embedded via automatic differentiation of the network output with respect to its inputs, computing exact partial derivatives at every training point.

### 2.2.2 Model Architecture

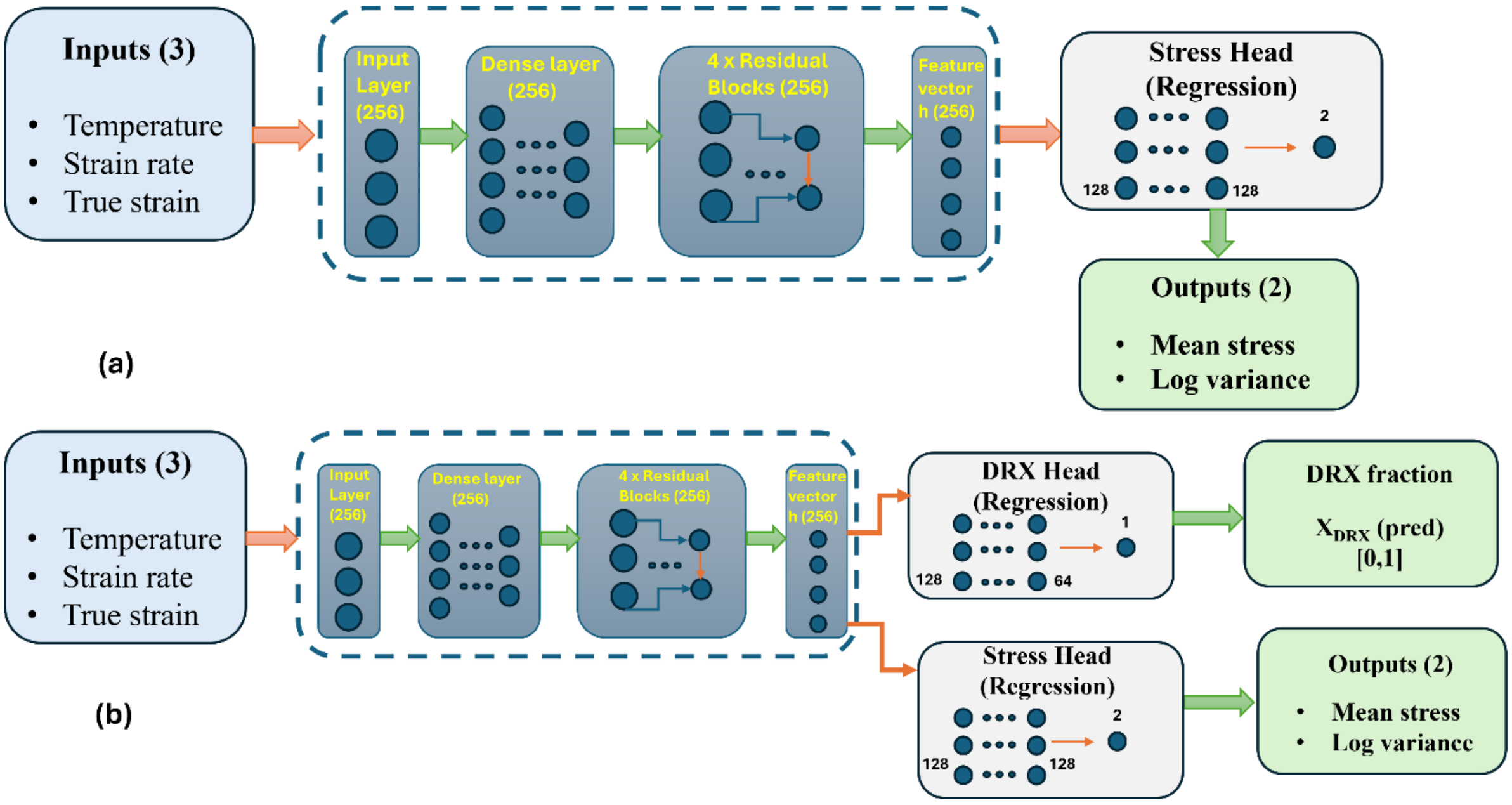

*Figure 1*: Schematic representation of the two STAR-PINN architectures developed for hot-deformation constitutive modelling. (a) Improved STAR-PINN (Model A), in which temperature, strain rate, and true strain are fed into a shared stacked residual encoder comprising an input layer, a dense feature-extraction layer, and four residual blocks. (b) DRX-Aware STAR-PINN (Model B), which retains the same residual encoder but introduces an additional DRX-regression head to predict the evolving DRX fraction, ($X_{DRX}$), alongside the stress-regression head.

#### 2.2.2.1 Improved STAR-PINN Architecture

The Improved Stacked Residual Physics-Informed Neural Network (STAR-PINN) was developed to capture the highly nonlinear hot-deformation response of the alloy. The network receives [$\varepsilon$, $\ln(\dot{\varepsilon})$, T] as inputs and propagates them through a sequence of residual blocks (Figure 1a). Each residual block contains fully connected layers, SiLU activation functions, layer normalization, dropout regularization, and a residual skip connection.

Residual learning is expressed as SiLU was chosen over ReLU for its everywhere-smooth differentiability, required by second-order physics penalties. Layer Normalization stabilizes training under the adaptive physics weight schedule. Dropout ($\tau = 0.05$) provides both regularization and MC-Dropout uncertainty at inference.

#### 2.2.2.2 Data Loss

Heteroscedastic Gaussian NLL:

$$L_{data} = \frac{1}{N}\sum [\ 0.5\cdot\exp(-\log \sigma^2)\cdot(\sigma - \mu)^2 + 0.5\cdot\log\sigma^2\ ] \quad (1.1)$$

where $\mu$ = predicted mean, $\log \sigma^2$ = predicted log-variance, N = number of training samples in the batch/set.

#### 2.2.2.3 Physics Loss

The total loss function combines prediction accuracy and physics regularization:

$$L_A = L_{data} + \lambda(t)\ (L_{thermal} + L_{SR} + 0.5L_{hard} + L_{soft} + 0.3L_{curv}) \quad (1.2)$$

where $\lambda(t)$ is an adaptive weighting parameter. The thermal softening term penalizes violations of $\frac{\partial\sigma}{\partial T} < 0$, the strain-rate term enforces $\frac{\partial\sigma}{\partial \ln(\dot{\varepsilon})} > 0$, the hardening term promotes positive work hardening before peak stress, and the softening term encourages stress reduction after recrystallization onset. A curvature regularization term was introduced to suppress nonphysical oscillations in the predicted flow curves.

#### 2.2.2.4 DRX-Aware STAR-PINN Architecture

Although the Improved STAR-PINN captures constitutive behavior accurately, it does not explicitly account for microstructural evolution. To address this limitation, DRX-Aware STAR-PINN architecture was developed. The framework retains the residual stacked backbone but introduces an auxiliary branch that predicts the dynamic recrystallization fraction ($X_{DRX}$). The predicted DRX fraction is concatenated with latent features before the final stress prediction layer, allowing microstructural evolution to directly influence constitutive response (Figure

1b).

The critical strain for DRX initiation was estimated from experimental flow curves using:

$$\varepsilon_c = 0.6\,\varepsilon_p \quad (1.3)$$

where $\varepsilon_p$ denotes the peak strain. For the present alloy, $\varepsilon_p$ = 0.1459 and $\varepsilon_c$ = 0.0876.

**2.2.2.5 Additional Physics Losses (DRX-Aware STAR-PINN model)**

| Term | Weight | Purpose |
|---|---|---|
| $L_{DRX}$ | 1.0 | DRX monotonicity |
| $L_{Avrami}$ | 1.0 | JMAK kinetic prior |
| $L_{couple}$ | 1.0 | DRX↑ → σ↓ coupling |
| $L_{sat}$ | 0.5 | $X_{DRX}$ → 1 at large ε |
| $L_{Arr}$ | 0.5 | Kinetic consistency |
| L_anchor | 0.05 | Regularises learnable JMAK (k, n) toward literature-motivated initial values |

*Table 2: List of physics losses in model B with their purpose.*

$$X_{JMAK} = 1 - \exp[-k\cdot((\varepsilon-\varepsilon_c)/(\varepsilon_p-\varepsilon_c))^n] \quad (1.4)$$

k and n are learnable parameters, reparameterised through a bounded sigmoid transform (k ∈ [0.2, 8.0], n ∈ [0.5, 4.0]) and initialised at $k_0$ = 2.0, $n_0$ = 2.0. A soft anchor loss (Eq. 1.5, weight $W_{anchor}$ = 0.05, applied without the λ(t) warm-up scaling so it is active from the first training step) regularises k and n toward these initial values, preventing the parameters from drifting to a degenerate solution while still allowing the network to adapt them to the data. Training converged to k = 2.0288, n = 1.9921 in the main run, and to k = 2.0138 ± 0.0139, n = 1.9935 ± 0.0020 averaged across the five cross-validation folds, in both cases within 2% of the literature-motivated initial values, indicating that the fixed k = n = 2.0 assumption used in earlier constitutive treatments is already an appropriate description of this alloy's DRX kinetics.

Total DRX-Aware STAR-PINN loss:

$$L_B = L_{data} + \lambda(t)\,(W_{thermal}\cdot L_{thermal} + W_{SR}\cdot L_{SR} + W_{hard}\cdot L_{hard} + W_{soft}\cdot L_{soft} + W_{curv}\cdot L_{curv}) + \lambda(t)\,(W_{DRX}\cdot L_{DRX} + W_{Avrami}\cdot L_{Avrami}) + W_{couple}\cdot L_{couple} + W_{sat}\cdot L_{sat} + W_{Arr}\cdot L_{Arr} + W_{anchor}\cdot L_{anchor}(k,n) \quad (1.5)$$

where $L_{phys}$ contains the thermal, strain-rate, hardening, and softening constraints; $L_{DRX}$ enforces recrystallization evolution; $L_{Avrami}$ promotes agreement with Avrami kinetics; $L_{couple}$

links DRX evolution with stress reduction; $L_{sat}$ enforces saturation behavior; and $L_{Arr}$ introduces Arrhenius consistency. $L_{anchor}$ is a soft prior pulling the learnable JMAK parameters back toward their literature-motivated initialisation; unlike the other physics terms, it is applied without λ(t) scaling so that it is active from the first training step.

| Loss term | Definition |
|---|---|
| $L_{thermal}$ | $\mathrm{mean}(\mathrm{ReLU}(\partial\sigma/\partial T))$ |
| $L_{SR}$ | $\mathrm{mean}(\mathrm{ReLU}(-\partial\sigma/\partial\dot{\varepsilon}))$ |
| $L_{hard}$ | $\mathrm{mean}(\mathbb{1}[\varepsilon<\varepsilon_c]\cdot\mathrm{ReLU}(-\partial\sigma/\partial\varepsilon))$ |
| $L_{soft}$ | $\mathrm{mean}(\mathbb{1}[\varepsilon\geq\varepsilon_c]\cdot\mathrm{ReLU}(\partial\sigma/\partial\varepsilon))$ |
| $L_{curv}$ | $\mathrm{mean}(\mathbb{1}[\varepsilon\geq\varepsilon_c]\cdot\mathrm{ReLU}(\partial^2\sigma/\partial\varepsilon^2))$ |
| $L_{DRX}$ | $\mathrm{mean}(\mathbb{1}[\varepsilon\geq\varepsilon_c]\cdot\mathrm{ReLU}(-\partial X_{DRX}/\partial\varepsilon))$ |
| $L_{Avrami}$ | $\mathrm{MSE}(X_{DRX}, X_{JMAK})$ |
| $L_{couple}$ | $\mathrm{mean}(\mathbb{1}[\varepsilon\geq\varepsilon_c]\cdot\mathrm{ReLU}(\partial X_{DRX}/\partial\varepsilon\cdot\partial\sigma/\partial\varepsilon))$ |
| $L_{sat}$ | $\mathrm{mean}(\mathbb{1}[\varepsilon\geq\varepsilon_p]\cdot(X_{DRX}-1)^2)$ |
| $L_{Arr}$ | $\mathrm{MSE}(\ln\dot{\varepsilon}_{pred}, \ln\dot{\varepsilon}_{exp})$, with the sinh-Arrhenius relation defining $\ln\dot{\varepsilon}_{pred}$ |
| $L_{anchor}$ | $((k-k_0)/(k_{max}-k_{min}))^2 + ((n-n_0)/(n_{max}-n_{min}))^2$ |

*Table 3. Definitions of loss terms used for both STAR-PINN models*

### 2.2.2.6 Model Training and Hyperparameters

Both models were implemented in PyTorch and trained with the following configuration:

| Hyperparameter | Value |
|---|---|
| Optimizer | AdamW |
| Learning rate | $1\times10^{-3}$ |
| Weight decay | $1\times10^{-5}$ |
| LR scheduler | Cosine annealing ($\eta_{min} = 1\times10^{-6}$) |
| Max epochs | 15,000 |
| Mini-batch size | 512 |
| Early stopping patience | 1,500 epochs (on validation NLL) |
| Gradient clipping | max_norm = 5.0 |
| Encoder width | 256 (4 residual blocks) |
| Dropout rate | $\tau = 0.05$ (also used for MC-Dropout UQ) |
| Physics λ warm-up | $1\times10^{-5} \rightarrow 1\times10^{-2}$ (linear) |
| Arrhenius warm-up | Active after epoch 3,000 (Model B only) |
| MC-Dropout passes | M = 50 (inference) |
| Cross-validation | 5-fold (fold epochs = 5,000) |
| Activation function | SiLU (encoder & heads) |
| Normalization | Min–Max (per feature, training set only) |
| Train / Test split | 80 % / 20 % (random, seed = 42) |

*Table 4. Complete hyperparameter configuration used for both STAR-PINN models.*

#### 2.2.2.7 Uncertainty Quantification and Evaluation Metrics

Monte Carlo Dropout: M = 50 stochastic forward passes with dropout ON at inference yield predictive mean $\hat{\mu}$ (x) and std $\hat{\sigma}$ (x). The 95 % confidence interval is $\hat{\mu} \pm 1.96\hat{\sigma}$.

Model performance was assessed using:

$$RMSE = \sqrt{\frac{1}{N}\sum(\sigma - \hat{\sigma})^2} \quad (1.6)$$

$$MAE = \frac{1}{N}\sum|\sigma - \hat{\sigma}| \quad (1.7)$$

$$R^2 = 1 - \sum\frac{(\sigma - \hat{\sigma})^2}{\sum(\sigma - \bar{\sigma})} \quad (1.8)$$

## 3. Results and Discussion3.1 Training Convergence

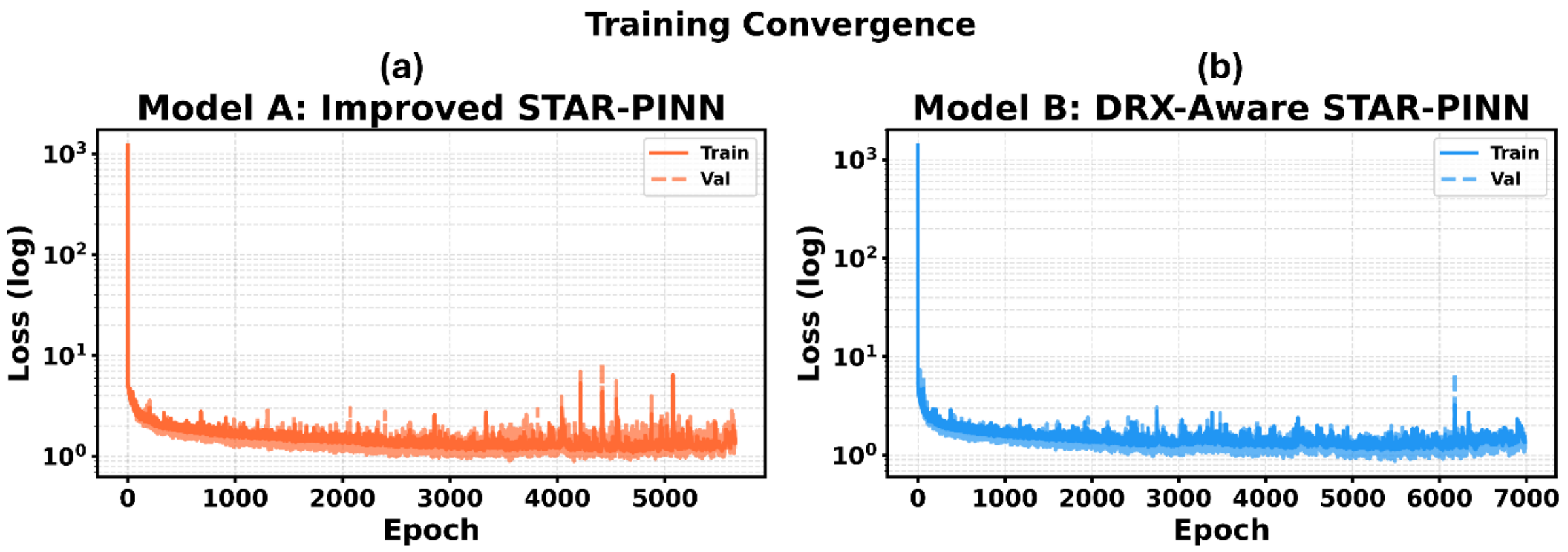

*Figure 2: The evolution of the total training loss (solid line) and validation loss (dashed line) on a logarithmic scale for both the (a) Improved STAR-PINN and (b) DRX-Aware STAR-PINN (blue) models.*

Both models exhibit characteristic two-stage convergence behaviour, consisting of a rapid initial reduction in loss followed by a prolonged refinement stage. During the initial stage, the loss decreases by approximately three orders of magnitude, from values of order $10^3$ to values of order $10^1$, within the first few hundred epochs. This is followed by a slower convergence regime in which the training and validation losses gradually decrease while exhibiting small stochastic fluctuations associated with mini-batch optimisation and the evaluation of physics-based residuals. The Improved STAR-PINN reached its early-stopping criterion at epoch 5653, whereas the DRX-Aware STAR-PINN continued to epoch 6985 before early stopping, reflecting the additional optimisation requirements associated with its DRX branch and coupled physics constraints. For the DRX-Aware STAR-PINN, the JMAK parameters were treated as learnable variables during training; they converged from the initial values k=2.0 and n=2.0 to k=2.0288 and n=1.9921, respectively, indicating only a small adjustment of the assumed recrystallisation kinetics during joint optimisation. Following the implementation of the numerical safeguards and the staged activation of the physics terms, no loss divergence or NaN events were observed during either training procedure. The close tracking of the training and validation losses throughout the latter part of training indicates that neither model exhibits pronounced overfitting under the adopted train–test partition. The fluctuations observed during the plateau region are characteristic of mini-batch optimisation in which physics residuals are evaluated over different subsets of the training data. Their overall magnitude remains bounded as the cosine-annealing learning-rate schedule progressively reduces the learning rate. These results demonstrate stable numerical convergence of both architectures, while the additional

DRX-aware formulation achieves convergence despite the presence of the coupled DRX, softening, Avrami, saturation, and Arrhenius-related constraints.

### 4.3 Predictive Performance on the Random Validation Subset

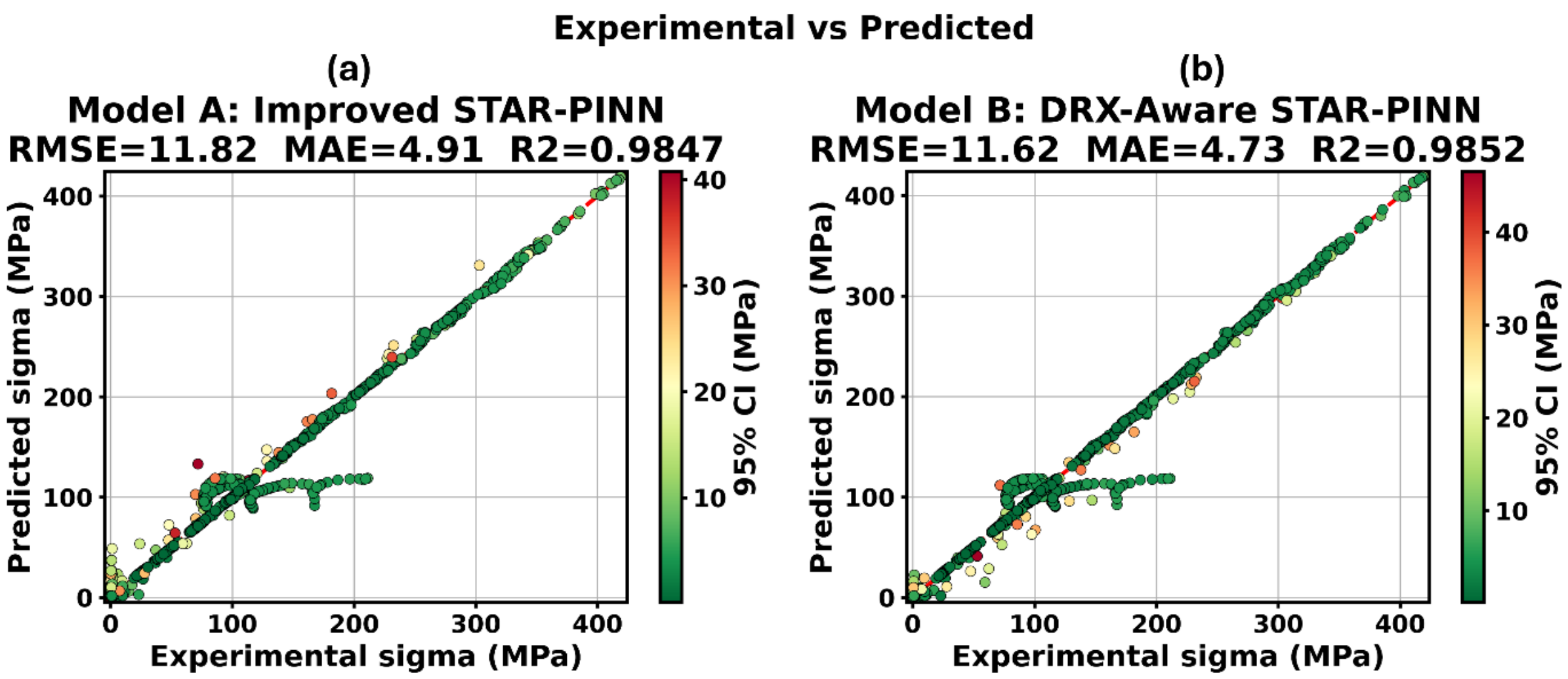

*Figure 3: The parity plots of experimental versus predicted flow stress for (a) Improved STAR-PINN and (b) DRX-Aware STAR-PINN with individual points color-coded by their MC-dropout 95% confidence interval half-width.*

| Model | RMSE (MPa) | MAE (MPa) | R2 | 95% CI (MPa)) |
|---|---|---|---|---|
| **Imp. STAR-PINN** | 11.82 | 4.91 | 0.9847 | ±3.12 (mean) |
| **DRX-Aware STAR-PINN** | 11.62 | 4.73 | 0.9852 | ±3.02 (mean) |

*Table 5. Validation-subset predictive performance metrics. $R^2$ = coefficient of determination; RMSE = root mean square error; MAE = mean absolute error; 95% CI = Monte Carlo dropout confidence interval half-width (mean across test set).*

Both models attain high predictive accuracy on the random validation subset, with the DRX-Aware STAR-PINN achieving a slightly higher $R^2$ of 0.9852 compared with 0.9847 for the Improved STAR-PINN. The corresponding RMSE and MAE are also marginally lower for the DRX-Aware model, decreasing from 11.82 to 11.62 MPa and from 4.91 to 4.73 MPa, respectively. Although these differences are relatively small, they indicate that incorporating the predicted DRX fraction into the stress head does not compromise predictive accuracy and provides a modest improvement in this random-partition evaluation. Examination of the parity plots shows that the largest prediction errors, represented by the orange-to-red points on the confidence-interval scale, are concentrated primarily in the intermediate stress range of approximately 100–150 MPa, with confidence-interval half-widths reaching approximately 40 MPa for individual observations. This region corresponds predominantly to the transition between strain hardening and post-peak softening at intermediate temperatures and strain rates, where small changes in thermomechanical conditions can produce substantial variations in flow stress. The DRX-Aware STAR-PINN shows somewhat fewer high-uncertainty points in this region, suggesting that explicit incorporation of the DRX state may improve the representation of the stress response during the hardening-to-softening transition. However, because this evaluation is based on a single random 80/20 partition, the observed difference should be interpreted as a modest performance advantage rather than definitive evidence of superior generalization. The robustness of the comparison is therefore further assessed using five-fold cross-validation in Section 4.6.

### 4.4 Flow Curve Reconstruction with Uncertainty Quantification

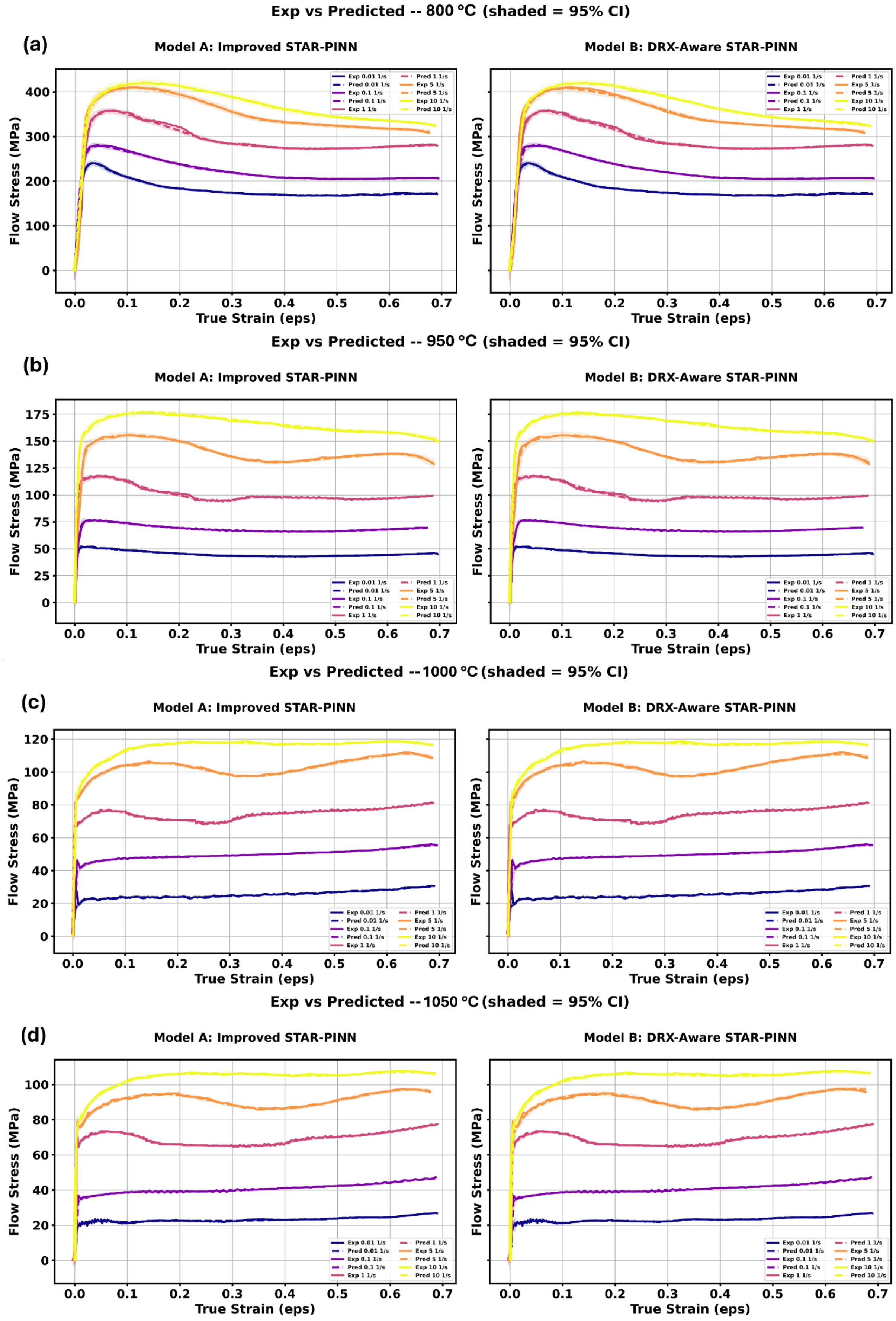

Exp vs Predicted -- 800 °C (shaded = 95% CI)
(a)
Model A: Improved STAR-PINN
Model B: DRX-Aware STAR-PINN
Exp vs Predicted -- 950 °C (shaded = 95% CI)
(b)
Model A: Improved STAR-PINN
Model B: DRX-Aware STAR-PINN
Exp vs Predicted -- 1000 °C (shaded = 95% CI)
(c)
Model A: Improved STAR-PINN
Model B: DRX-Aware STAR-PINN
Exp vs Predicted -- 1050 °C (shaded = 95% CI)
(d)
Model A: Improved STAR-PINN
Model B: DRX-Aware STAR-PINN
Flow Stress (MPa)
True Strain (eps)
Exp 0.01 1/s
Pred 0.01 1/s
Exp 0.1 1/s
Pred 0.1 1/s
Exp 1 1/s
Pred 1 1/s
Exp 5 1/s
Pred 5 1/s
Exp 10 1/s
Pred 10 1/s

*Figure 4: The reconstruction of complete true stress–true strain flow curves at all experimental temperatures (a) 800 °C, (b) 950 °C, (c)1000 °C, and (d)1050 °C and all five strain rates (0.01, 0.1, 1, 5, and 10 $s^{-1}$). Experimental data are shown as solid lines and model predictions as dashed lines of identical color, with the 95% MC-dropout confidence interval shaded in the corresponding hue.*

At 800 °C (Fig. 4a), the flow curves exhibit strain hardening followed by a clear peak and softening, with the highest stress approaching ~410 MPa at $\dot{\varepsilon}$ = 10 $s^{-1}$. The magnitude of the peak decreases systematically with decreasing strain rate, while the post-peak response gradually approaches a near-steady state. Both models reproduce the experimental curves closely across all five strain rates, including the peak position and the subsequent softening trajectory. The 95% CI bands remain narrow over most of the deformation range, with somewhat greater widths around the rapid hardening and peak-stress transition. The two architectures produce nearly overlapping predictions, indicating that the additional DRX-aware formulation does not substantially alter the overall stress response at this temperature.

At 950 °C (Fig. 4b), the flow stresses are substantially lower, with the highest peak reaching approximately 175 MPa at 10 $s^{-1}$. The curves still exhibit a distinct peak at the higher strain rates, whereas the lower-rate responses show weaker hardening and more gradual post-peak evolution. Both models closely follow the experimental trends throughout the strain range. The largest visible differences between prediction and experiment occur in the intermediate and post-peak regions, particularly for the lower and intermediate strain-rate curves, where the experimental response contains broader changes in slope. These differences are consistent with the difficulty of representing condition-dependent peak and softening behavior using a single global critical-strain reference within the physics formulation. Nevertheless, the predicted trajectories remain physically consistent and the uncertainty bands are generally narrow.

At 1000 °C and 1050 °C (Fig. 4c and 4d), the flow-stress level decreases further and the curves become progressively flatter, particularly at the lower strain rates. At 1000 °C, the 0.01 and 0.1 $s^{-1}$ responses show mainly continuous hardening followed by a broad plateau, whereas the 5 and 10 $s^{-1}$ curves retain more pronounced peak-and-softening characteristics. At 1050 °C, the same rate dependence is evident, although the peak becomes less pronounced and the flow curves approach stable stress levels after the initial hardening stage. Both STAR-PINN models capture these changes in curve morphology and reproduce the experimental responses with closely overlapping predictions. The 95% CI bands remain comparatively narrow across both temperatures, including the regions where the transition from hardening to plateau behavior is gradual. The close agreement between the two models indicates that the DRX-aware formulation preserves the constitutive trends captured by the Improved STAR-PINN while providing a clear mechanism-informed representation of microstructural evolution. Overall, Figure 4 demonstrates that both architectures can represent the strong temperature- and strain-rate-dependent transition from high-stress peak-and-softening behavior at 800 °C toward lower-stress, more recovery-dominated responses at 1000–1050 °C.

### 4.5 Gradient-Based Feature Importance

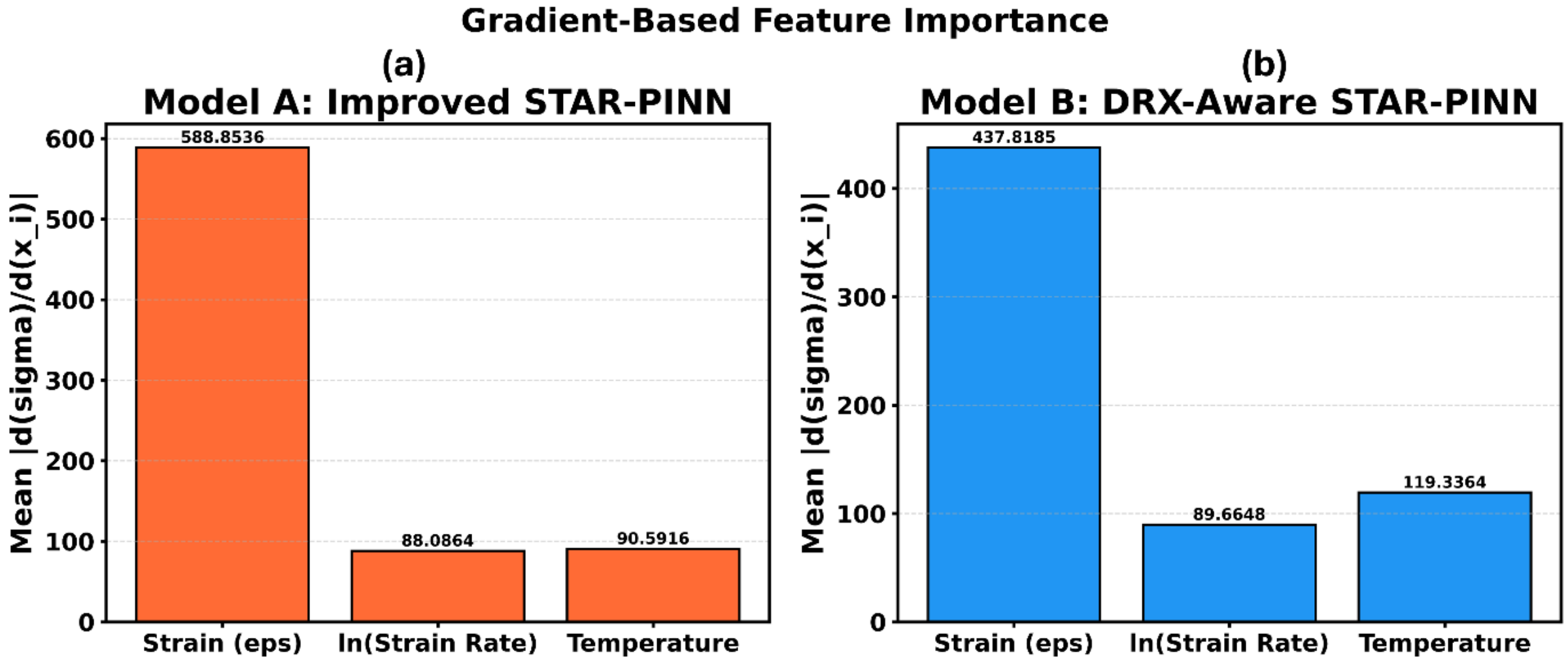

*Figure 5: The gradient-based feature importance, (a) Improved STAR-PINN and (b) DRX-Aware STAR-PINN, defined as the mean absolute gradient of predicted stress with respect to each normalized input feature, evaluated over the full test set.*

Multiple significant insights arise from the gradient-based feature-importance analysis. True strain is the primary input in both models, evidenced by mean absolute gradient importances of 588.85 for the Improved STAR-PINN and 437.82 for the DRX-Aware STAR-PINN. In Model A, the significance of strain is about 6.7 times greater than that of ln(strain rate) (88.09) and 6.5 times greater than temperature (90.59), emphasizing the primary influence of accumulated strain in controlling the development of flow stress across the examined deformation range. This aligns with the noted shift from initial deformation via strain hardening, peak stress, and the later flow softening linked with DRX.

$$Z = \dot{\varepsilon} \cdot \exp(Q/RT) \qquad (1.9)$$

The introduction of the DRX branch produces a clear redistribution of input sensitivity. The strain importance decreases from 588.85 to 437.82, corresponding to a reduction of approximately 25.7%, while the importance of $\ln(\dot{\varepsilon})$ changes only marginally, from 88.09 to 89.66 (approximately 1.8% increase). In contrast, the temperature importance increases from 90.59 to 119.34, representing an increase of approximately 31.8%. Thus, the DRX-Aware STAR-PINN does not simply reduce the overall sensitivity to the original inputs; rather, it redistributes the constitutive sensitivity away from direct strain dependence and toward temperature and strain-rate-related effects. This redistribution carries physical significance since the DRX-aware formulation integrates $X_{DRX}$ as a middle representation that clearly represents the progression of recrystallization in relation to strain, temperature, and strain rate. As a result, certain data that Model A needs to derive from the strain input is conveyed through the DRX pathway in Model B. This allows the stress head to depend less on the raw strain variable while becoming more attuned to the thermomechanical factors influencing DRX kinetics. The significant rise in temperature relevance aligns with the pronounced temperature sensitivity of thermally activated deformation and recrystallization rates. Importantly, the feature-importance results should not be interpreted as evidence that strain becomes less physically important in the DRX-Aware model. Strain remains overwhelmingly the most influential individual input (437.82), while the redistribution indicates that its effect is increasingly mediated through the learned DRX state. The results therefore provide evidence of a more distributed representation of the constitutive response in the DRX-Aware STAR-PINN, rather than simply an increase or decrease in overall model sensitivity.

### 4.6 Five-Fold Cross-Validation

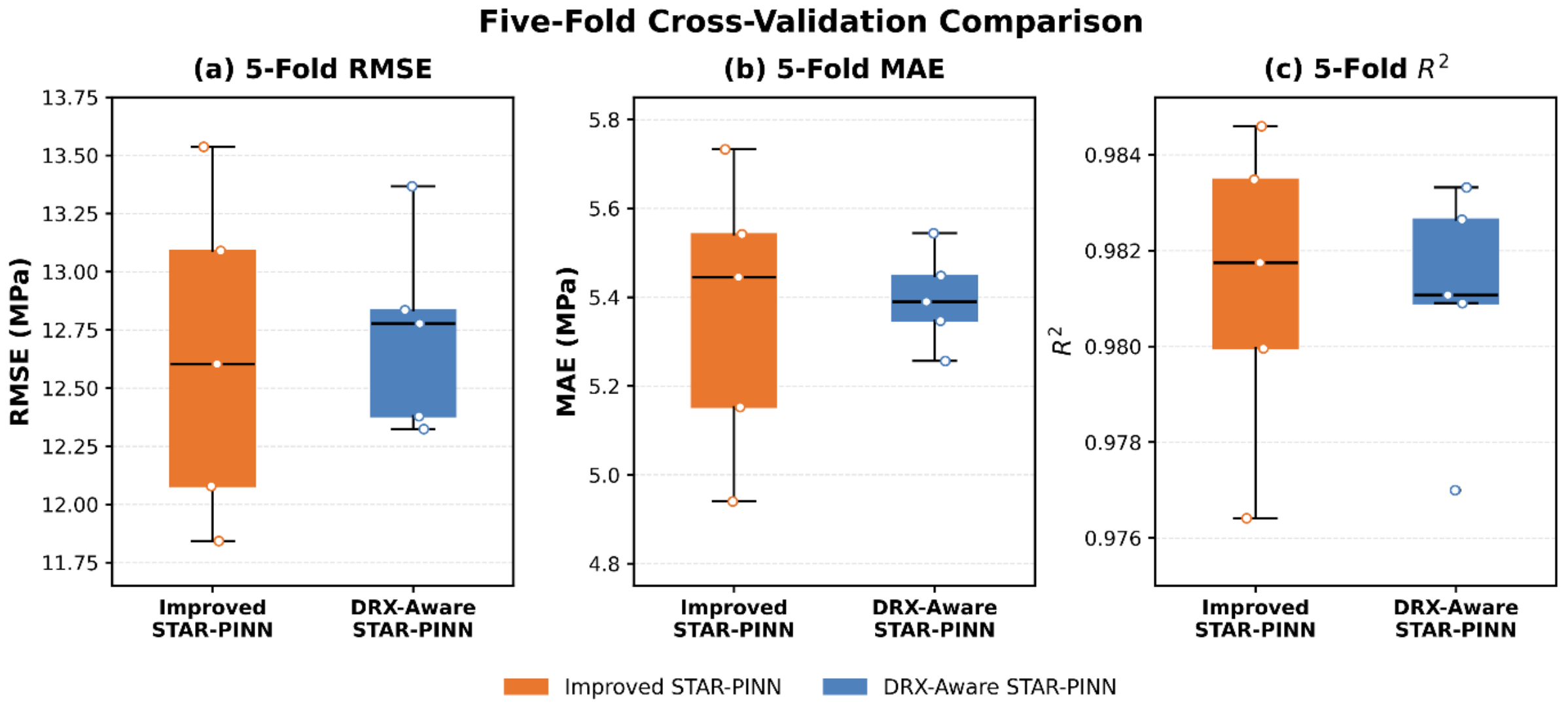

*Figure 6: The 5-fold cross-validation box plots for (a)RMSE, (b)MAE, and (c)$R^2$.*

| Model | CV RMSE (MPa) | CV MAE (MPa) | CV $R^2$ |
|---|---|---|---|
| **Imp. STAR-PINN** | 12.63 ± 0.63 | 5.36 ± 0.32 | 0.9812 ± 0.0029 |
| **DRX-Aware STAR-PINN** | 12.74 ± 0.38 | 5.40 ± 0.11 | 0.9810 ± 0.0022 |

*Table 6. Five-fold cross-validation summary. Values reported as mean ± standard deviation across five folds.*

The results from five-fold cross-validation further demonstrate the reliability of both STAR-PINN formulations. The Improved STAR-PINN demonstrated a median RMSE of 12.60 MPa, whereas the DRX-Aware STAR-PINN revealed a marginally elevated median RMSE of 12.78 MPa, with considerable overlap in their interquartile ranges. The median MAE values were 5.45 and 5.39 MPa, correspondingly, whereas the median $R^2$ values were 0.9818 and 0.9811 for the Improved and DRX-Aware STAR-PINNs, respectively. Consequently, both architectures demonstrate almost identical interpolation performance over the five folds, even with the extra DRX-related physics included in Model B. Importantly, the DRX-Aware STAR-PINN showed reduced fold-to-fold variability compared to the Improved STAR-PINN, with standard deviations of 0.38 MPa against 0.63 MPa for RMSE, 0.11 MPa versus 0.32 MPa for MAE, and 0.0022 versus 0.0029 for $R^2$, respectively. The Improved STAR-PINN showed increased variability in RMSE, mainly linked to the higher-error folds (notably folds 3 and 5), while the DRX-Aware formulation stayed more closely grouped around its average performance. Crucially, since the cross-validation method employs randomly divided samples from the identical experimental domain, these findings illustrate resilience to the partitioning of training data and interpolation within the sampled processing window; however, they ought not to be seen as proof of out-of-distribution generalization. The CV outcomes indicate that the DRX-Aware STAR-PINN achieves predictive accuracy similar to the more straightforward Improved STAR-PINN, while showing reduced fold-to-fold variability and offering the extra functionality of physically interpretable DRX predictions.

### 4.7 Uncertainty Assessment and Quantile–Quantile Analysis of Prediction Errors

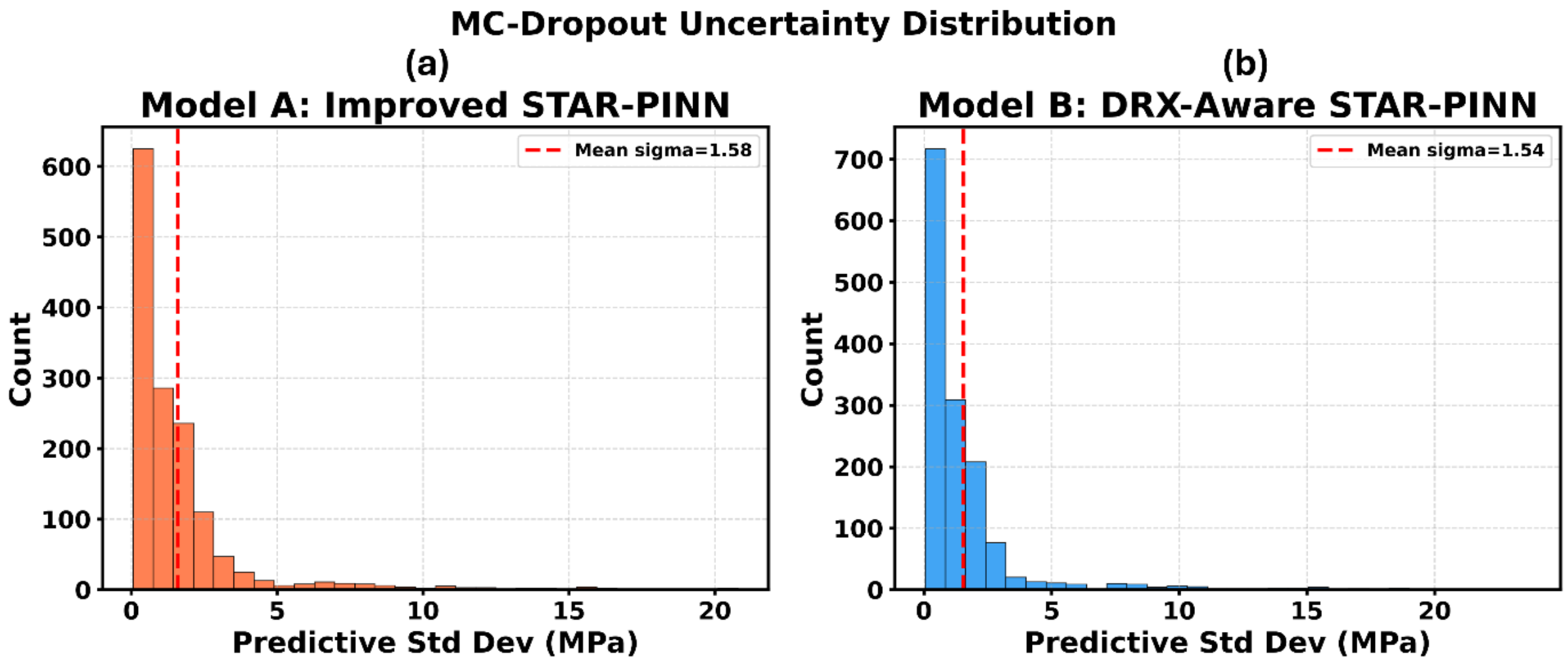

*Figure 7: The distribution of MC-dropout predictive standard deviations $\hat{\sigma}$ over the test set for (a) Improved STAR-PINN and (b) DRX-Aware STAR-PINN models. Both distributions are strongly right skewed: the overwhelming majority of test points (>85%) have predictive standard deviations below 2 MPa, with long tails extending to ~20 MPa.*

The Improved STAR-PINN achieves a mean predictive standard deviation of 1.59 MPa, corresponding to a mean 95% confidence interval of approximately ±3.12 MPa. The DRX-Aware STAR-PINN exhibits a slightly higher mean predictive standard deviation of 1.70 MPa, corresponding to a mean 95% confidence interval of approximately ±3.35 MPa. The modest increase in predictive uncertainty for the DRX-Aware model is consistent with its more complex architecture, which introduces an additional DRX prediction pathway coupled to the stress response. During MC-dropout inference, stochastic variations in the shared feature representation can therefore influence both the predicted DRX fraction, $X_{DRX}$, and the subsequent stress prediction. This additional coupled pathway provides a plausible explanation for the slightly broader uncertainty distribution observed for the DRX-Aware STAR-PINN. Physically, the largest predictive uncertainties, represented by the right tail of the distribution (>10 MPa), are expected to occur in deformation regimes where the constitutive response is particularly sensitive to the processing conditions. In particular, the peak-stress transition region (ε≈0.05–0.15), especially at high strain rates and lower temperatures, represents a challenging region for the model because small variations in temperature, strain rate, and accumulated strain can produce appreciable changes in the predicted stress and DRX evolution. The concentration of higher uncertainty in this transition regime therefore provides evidence that the MC-dropout estimate captures regions of comparatively lower model confidence rather than behaving as a purely random error indicator. Importantly, this interpretation should be regarded as an indirect validation of the uncertainty estimate, rather than a direct experimental validation of predictive uncertainty.

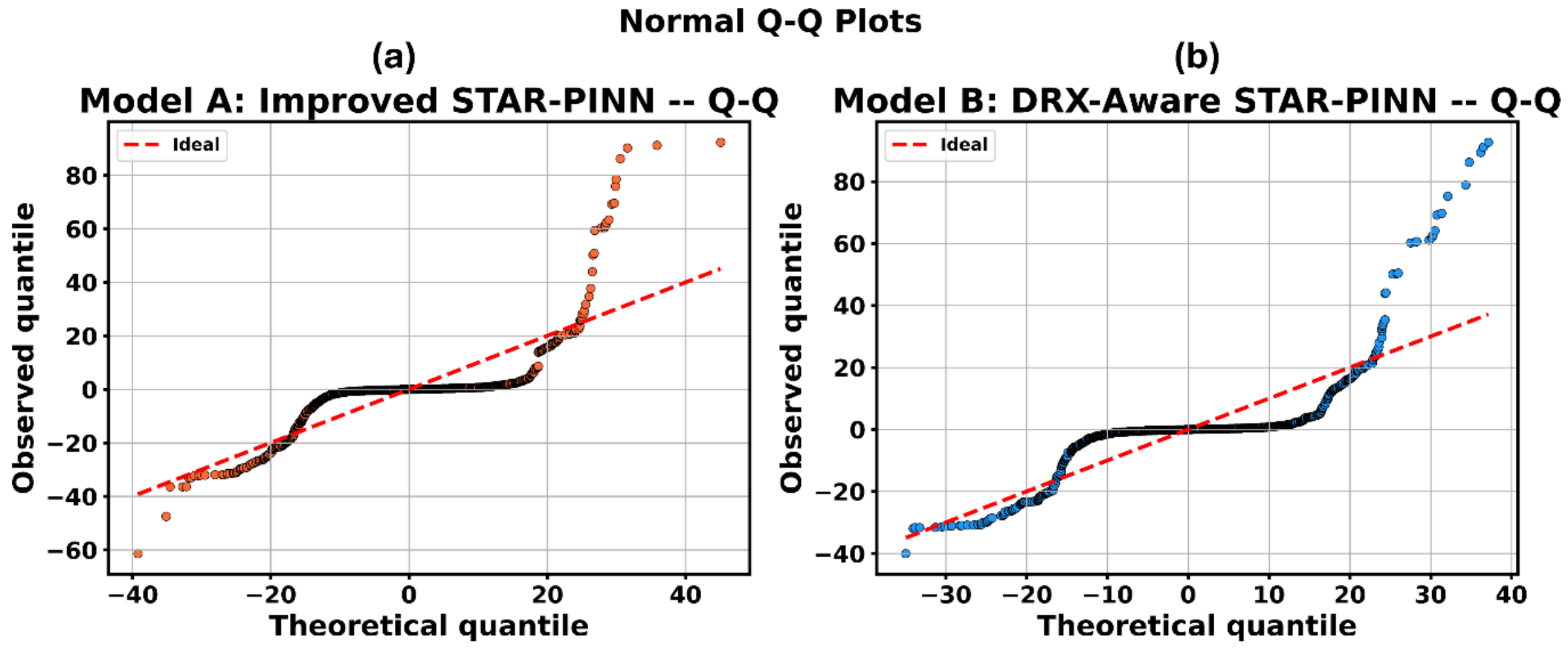

*Figure 8: The Q–Q plots of residuals for the (a) Improved STAR-PINN and (b) DRX-Aware STAR-PINN models.*

Deviations from this line suggest departures from the norm and could uncover prediction errors or significant outliers. In both models, the main part of the residual distribution stays fairly near the ideal trend, showing that most predictions exhibit minor errors. This indicates that both frameworks reflect the primary constitutive response under the deformation conditions. However, deviations from linearity are noted at both extremes, indicating that the residuals do not follow a Gaussian distribution and that greater errors arise in areas of nonlinear constitutive behavior. Model A (Figure 8a) shows significant discrepancies in both tails, with measured quantiles reaching around −60 MPa on the lower side and exceeding 90 MPa on the upper side.

The upward bend at high positive quantiles signifies substantial positive residuals, potentially linked to errors related to the shift between strain hardening and flow softening. Model B (Figure 8b) displays a comparable central distribution but diminished negative-tail excursions, achieving −40 MPa, while its positive tail still reaches almost 90 MPa. Consequently, the enhancement in residual behavior is clearer in the lower tail compared to the upper tail, instead of indicating a return to normality. The leftover tail discrepancies show that nonlinear deformation areas continue to pose difficulties for both models. The narrower negative tail of the DRX-Aware model aligns with the physics constraints related to DRX, offering direction for depicting recrystallization and its interaction with post-peak softening. In general, the Q-Q analysis indicates improved residual behavior in certain areas for the DRX-Aware model, while showing that physics-informed constraints do not eradicate all localized prediction inaccuracies.

**4.8 Dynamic Recrystallization Fraction Prediction**

Prior to delving into the detailed analysis of the anticipated DRX fraction curves and contour map, a significant qualification is warranted. The DRX fraction outputs from the DRX-Aware STAR-PINN provide qualitative predictions meant to illustrate physically consistent microstructural trends, not precise measurements of the recrystallized volume fraction. The model was trained solely on macroscopic flow stress data; no direct microstructural measurements, such as EBSD-derived recrystallized area fractions, optical metallography grain counts, or X-ray line-broadening dislocation density estimates were available for training targets or validation reference points. As a result, the absolute values of $X_{DRX}$ predicted by the network contain intrinsic uncertainty that cannot be measured without separate microstructural validation. The JMAK parameters were treated as learnable rather than fixed, allowing the network to adapt them to the data within a physically bounded range ($k \in [0.2, 8.0]$, $n \in [0.5, 4.0]$; see Section 2.2.5) rather than relying solely on literature-typical values. Training converged to $k = 2.0288$, $n = 1.9921$ in the main run and to $k = 2.0138 \pm 0.0139$, $n = 1.9935 \pm 0.0020$ across five cross-validation folds, remaining within 2% of the initial $k = n = 2.0$ in every case. This directly tests, rather than assumes, the appropriateness of the fixed-parameter choice used in earlier constitutive treatments and confirms it is a reasonable description of this alloy's DRX kinetics rather than an arbitrary convention. The data shown here should thus be viewed as proof that the DRX-Aware STAR-PINN has acquired physically consistent microstructural patterns, accurate monotonicity, appropriate temperature and strain-rate sequencing, and correct saturation behavior, instead of as precise forecasts of experimental DRX fractions.

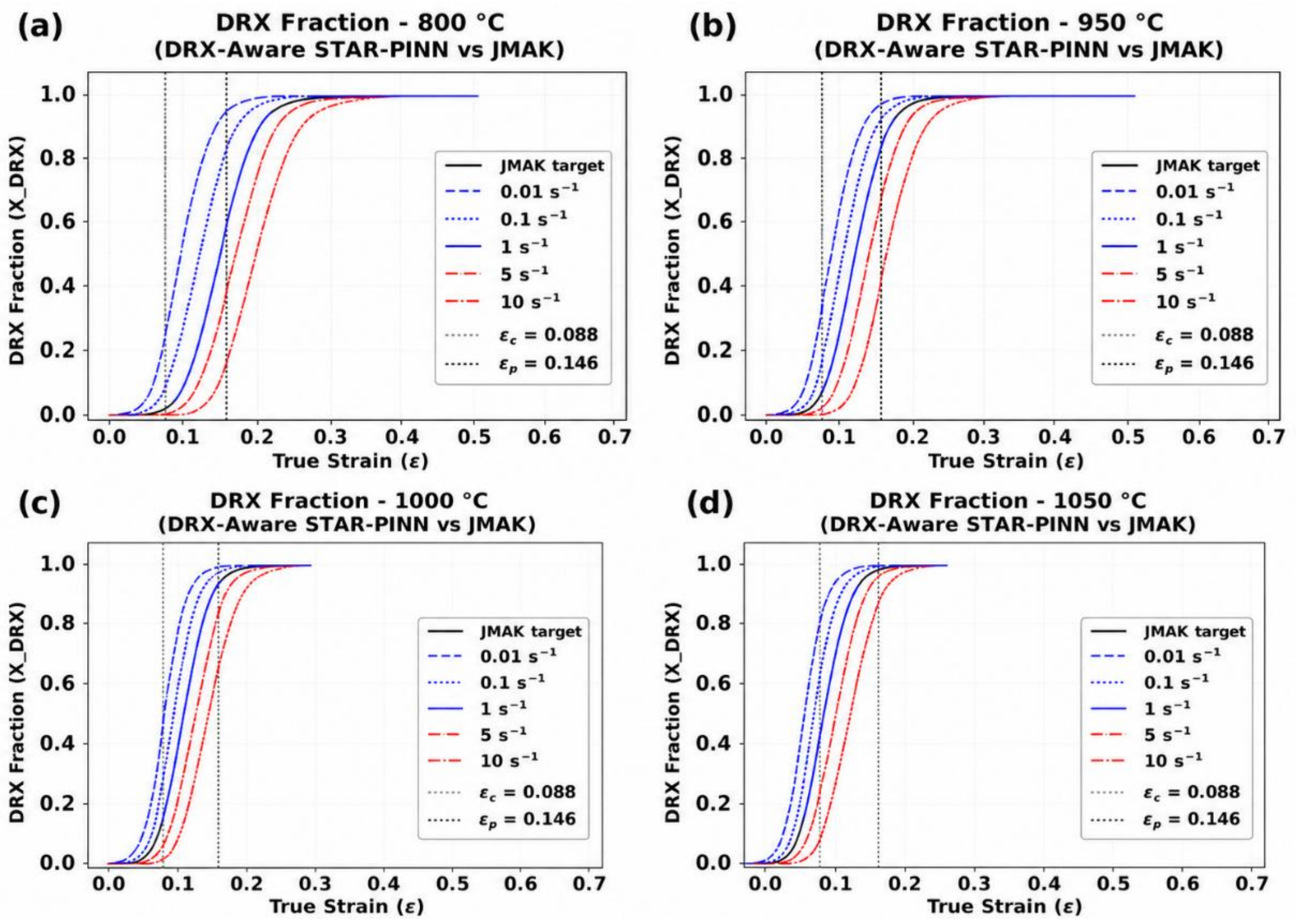

*Figure 9: The DRX fraction ($X_{DRX}$) as a function of true strain predicted by the DRX-Aware STAR-PINN at all four temperatures (a) 800°C, (b) 950°C, (c) 1000°C, (d) 1050°C and five strain rates, compared against the JMAK reference curve (thick black line) computed with the learned parameters (k = 2.0288, n = 1.9921). The vertical dotted lines indicate the critical strain $\varepsilon_c$ = 0.088 and peak strain $\varepsilon_p$ = 0.146.*

At each of the four temperatures, the anticipated XDRX curves show the typical sigmoidal pattern, staying near zero until the critical strain ($\varepsilon_c$ = 0.088), rising swiftly during the recrystallization phase, and nearing saturation ($X_{DRX} \rightarrow 1$) at higher strains. The strain-rate effect is now more distinctly clarified than in the earlier depiction. At 800 °C, raising the strain rate from 0.01 to 10 $s^{-1}$ gradually moves the forecasted DRX development to greater strains, suggesting postponed recrystallization with increasing deformation rates. This distinction remains clear at 950, 1000, and 1050 °C, though the curves start to converge more as the temperature rises. Consequently, the curves for higher strain rates reliably indicate that a specific DRX fraction is achieved later, while the lowest strain-rate condition achieves saturation at relatively lower strain levels. Throughout all temperature ranges, the forecasted curves largely align with the black JMAK reference, while permitting condition-specific variations via the network. This behavior suggests that the JMAK–Avrami regularization offers the general sigmoidal kinetic prior, while the trained network remains responsive to temperature and strain rate. The obtained parameters, k = 2.0288 and n = 1.9921, stay remarkably near the initial values of 2.0, reinforcing the appropriateness of the JMAK formulation as a zeroth-order representation of DRX kinetics for the studied alloy.

| **Condition** | **GOS-derived $X_{DRX}$ (2°)** | **Mean GOS (°)** | **Model-predicted $X_{DRX}$ (final ε)** |
|---|---|---|---|
| 1050 °C, 0.01 $s^{-1}$ | 0.86 | 0.6 | ≈ 1.0 |
| 1050 °C, 10 $s^{-1}$ | 0.35 | 1.5 | ≈ 1.0 |

Table 6. Comparison of GOS-derived DRX fraction and mean grain orientation spread (GOS) with the final DRX fraction predicted by the DRX-Aware STAR-PINN under selected deformation conditions.

A preliminary quantitative cross-check of the qualitative DRX kinetics mentioned earlier was conducted by quantifying the GOS-based recrystallized fraction using the available EBSD maps under two conditions (1050 °C, at 0.01 and 10 $s^{-1}$) [22]. Areas with GOS below a limit of 2° (also evaluated at 1° and 5° for sensitivity) were designated as recrystallized, and the calculated recrystallized area fraction was compared with the related DRX-Aware STAR-PINN prediction at the final experimental strain ($\varepsilon \approx 0.7$) for each corresponding condition. The agreement between the recrystallized fraction derived from GOS and that predicted by the model is only accurate at 1050 °C / 0.01 $s^{-1}$, where conditions are most thermodynamically favorable for significant DRX. At 1050 °C / 10 $s^{-1}$, the model overestimates the recrystallized fraction. This difference aligns with a particular, recognizable constraint of the current model: the JMAK saturation and Avrami regularization components are based on a singular, globally estimated critical strain $\varepsilon_c$ and peak strain $\varepsilon_p$ (Section 2.2.2.5), leading the expected DRX fraction to reach saturation at a constant strain, irrespective of strain rate. This comparison should be viewed as an initial, exploratory validation. A formulation of $\varepsilon_c$ and $\varepsilon_p$ that depends on conditions (such as temperature or Zener–Hollomon-parameter) or a completely learnable saturation constraint for each condition is recognized as a particular approach for enhancing the quantitative microstructural fidelity of DRX-Aware STAR-PINN in upcoming research.

### 4.9 Sensitivity of Predictive Accuracy to Physics-Loss Weights

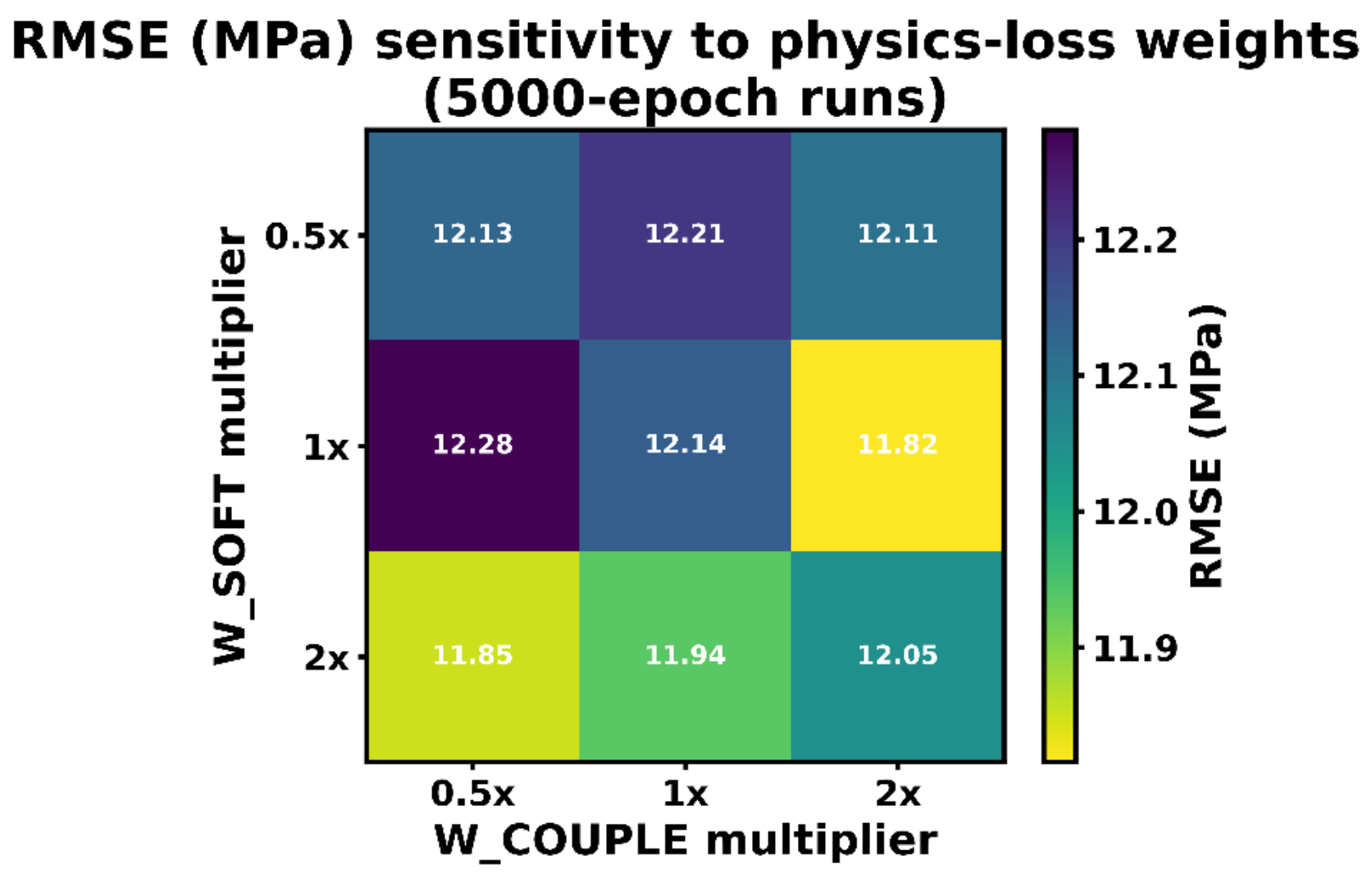

*Figure 10: Sensitivity of the DRX-Aware STAR-PINN prediction to the relative weighting of the post-peak softening and DRX–softening coupling physics losses. The heatmap shows the RMSE obtained for combinations of Wsoft and Wcouple equal to 0.5×, 1×, and 2× their baseline values, using a 5,000-epoch training budget for each configuration.*

To assess the sensitivity of the DRX-Aware STAR-PINN to the relative weighting of the physics constraints governing post-peak deformation, a systematic weight-sensitivity analysis was performed for the two most directly related terms, namely the post-peak softening loss ($W_{soft}$) and the DRX–softening coupling loss ($W_{couple}$). Each weight varied independently over 0.5×, 1×, and 2× its baseline value, resulting in nine combinations. All configurations were trained for 5,000 epochs using the same train–test partition, and their performance was evaluated on the held-out test set. As shown in Fig. 10, the RMSE varied from 11.82 to 12.28 MPa across the entire 3×3 parameter space, corresponding to a variation of only approximately 3.8% relative to the mean RMSE. The corresponding $R^2$ values varied between 0.9834 and 0.9847, indicating that the predictive performance remained relatively stable over the investigated range of physics-loss weights. The lowest RMSE in the sensitivity sweep, 11.82 MPa, was obtained for $W_{soft}$=1× and $W_{couple}$=2×. This result suggests that moderately increasing the weight assigned to the DRX–softening coupling constraint can improve stress prediction within the tested parameter range. In contrast, increasing $W_{soft}$ alone did not produce a monotonic improvement in accuracy. For example, at $W_{couple}$ =1×, changing $W_{soft}$ from 0.5× to 1× and 2× resulted in RMSE values of 12.21, 12.14, and 11.94 MPa, respectively, whereas at $W_{couple}$=2×, the corresponding values were 12.11, 11.82, and 12.05 MPa. This non-monotonic response indicates that the two constraints are coupled during optimization and that excessive emphasis on either term does not necessarily improve predictive accuracy. Overall, the relatively narrow variation in RMSE and $R^2$ across the investigated weight space indicates that the DRX-Aware STAR-PINN is not strongly dependent on the precise choice of these two physics-loss weights. The sensitivity analysis therefore supports the robustness of the adopted physics-informed formulation rather than suggesting that its performance arises from a narrowly tuned combination of penalty weights. Nevertheless, because these calculations were performed using a reduced 5,000-epoch training budget, the results should be interpreted as a local sensitivity analysis around the baseline configuration rather than as a global optimization of the loss weights. The slightly lower RMSE obtained at Wsoft=1× and Wcouple =2× suggests that adaptive or violation-based weighting could be explored in future work.

### 4.10 DRX Contour Map and 3D Constitutive Surface

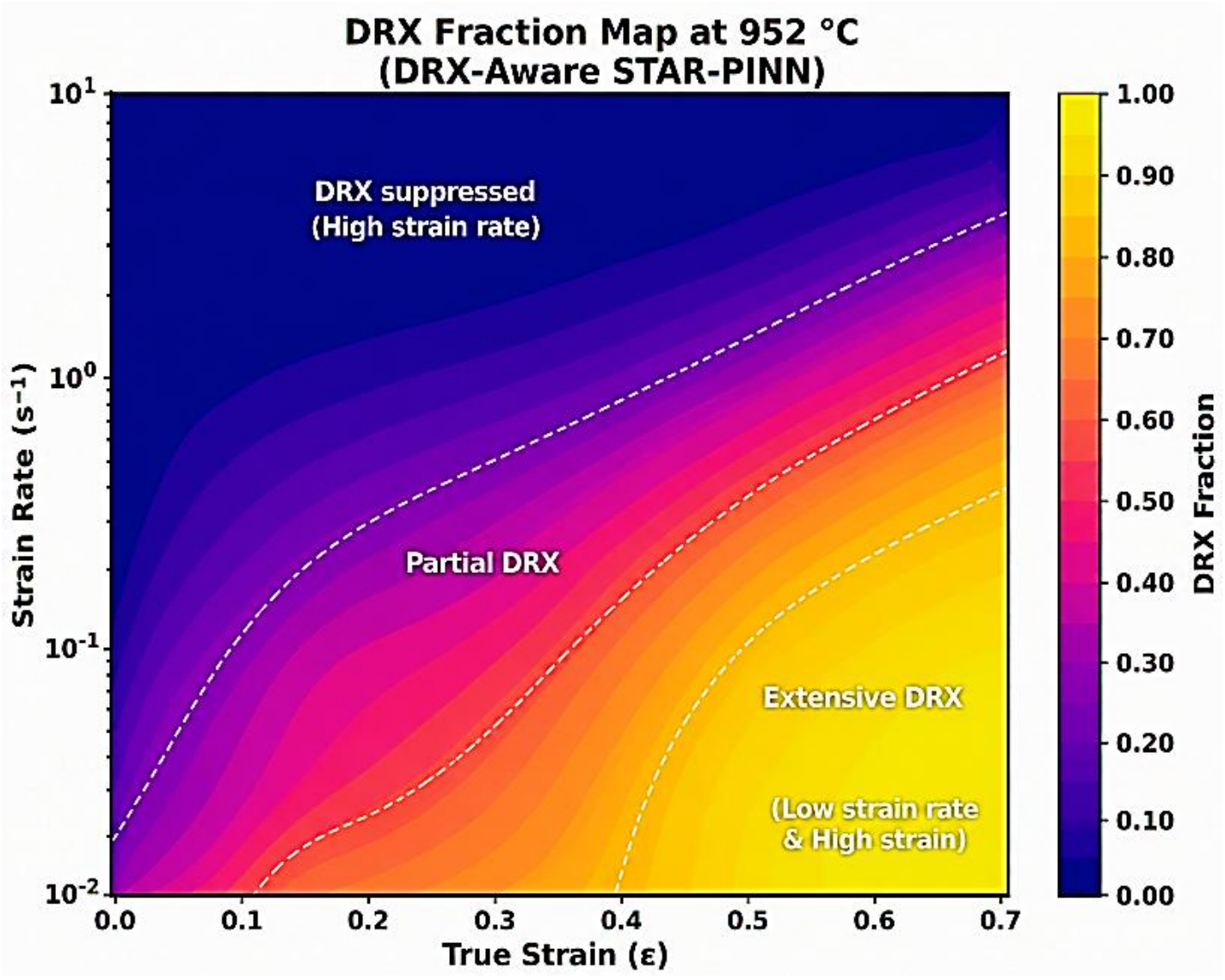

*Figure 11: The DRX fraction as a two-dimensional contour map over the ε–$\dot{\varepsilon}$ plane at the median temperature (~952 °C). The color scale runs from deep blue ($X_{DRX}$ = 0, pre-DRX regime) through orange and yellow ($X_{DRX} \rightarrow 1$, fully recrystallized).*

In Figure 11, the contour map at 952 °C illustrates the relationship between the predicted DRX fraction and both true strain and strain rate. At low strain rates, XDRX rises with greater strain, resulting in a high-DRX area at larger strains. With the increase in strain rate, the shift toward greater DRX fractions moves to higher strains, suggesting that recrystallization is delayed during quicker deformation. This pattern aligns with the anticipated decrease in time allotted for recrystallization processes at elevated strain rates. The map indicates that the transition isn't limited to one specific strain value; rather, the DRX evolution creates a wide, curved boundary throughout the strain-rate spectrum. At roughly ε = 0.1–0.2, the map shows significant DRX at reduced strain rates, while greater strains are necessary to achieve similar fractions at elevated strain rates. The yellow area associated with $X_{DRX}$ nearing unity is mainly focused on the high-strain and low-strain-rate sections of the map, rather than being evenly distributed across all conditions. The seamless transitions represent a gradual change in the anticipated recrystallized fraction, free from sudden numerical jumps. The white dashed lines depict the gradual progression of the DRX transition with strain rate and serve as a depiction of intermediate DRX conditions. In general, the anticipated map represents the competition between accumulated strain and deformation rate: higher strain encourages DRX, while a higher strain rate hinders or postpones its emergence. These trends reinforce the physical consistency of the DRX-aware formulation throughout the examined temperature and deformation range.

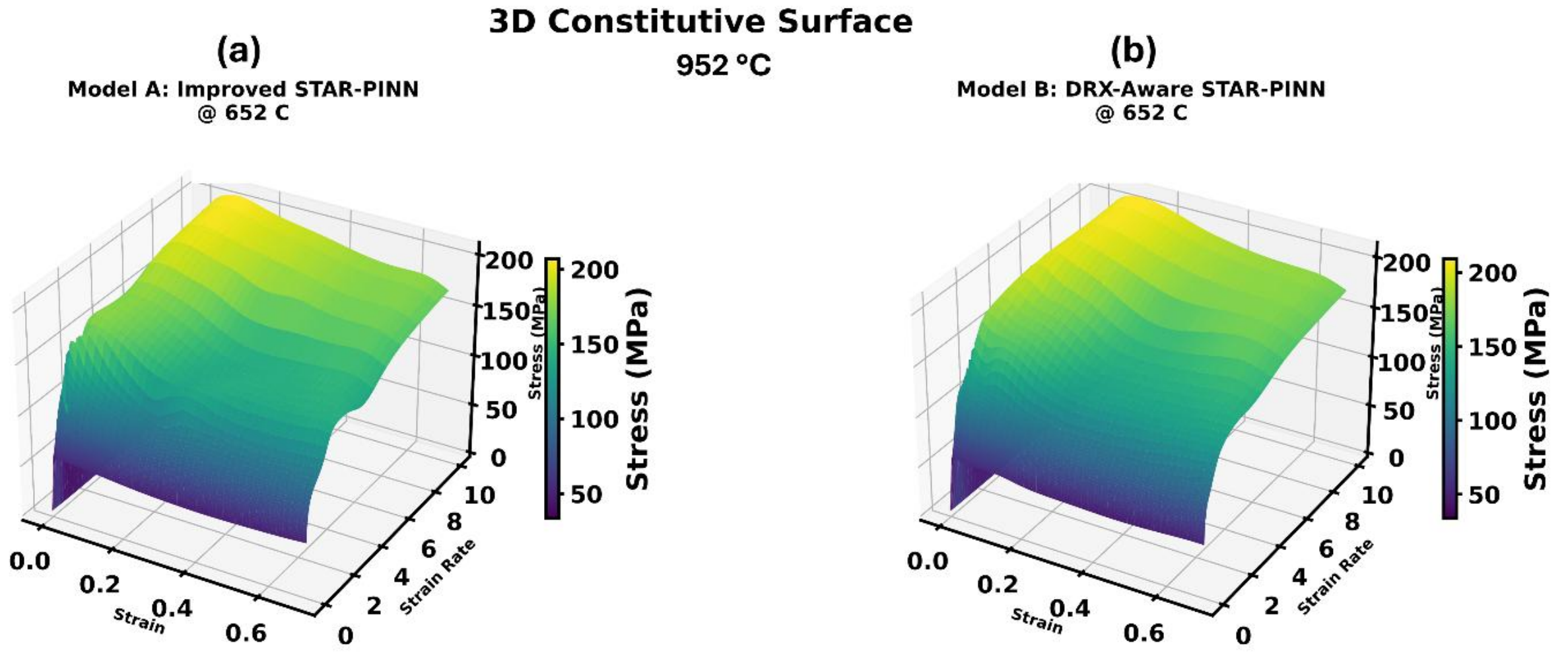

*Figure 12: Three-dimensional constitutive surfaces of predicted flow stress over the ε–ε̇ plane at the same reference temperature (~952 °C) for (a) Improved STAR-PINN and (b) DRX-Aware STAR-PINN models.*

From Figure 12, both surfaces exhibit a consistent qualitative topology, with flow stress increasing with strain rate and showing a distinct peak followed by post-peak softening as strain increases. At lower strain rates, the response becomes comparatively flatter, indicating reduced sensitivity to accumulated strain. The Improved STAR-PINN and DRX-Aware STAR-PINN surfaces are closely similar overall, with both reproducing the characteristic hardening–softening behavior across the investigated strain-rate range. The DRX-Aware formulation shows a slightly more regular post-peak response, consistent with the additional DRX-related constraints and coupling between recrystallization evolution and stress softening. Thus, both models maintain physically plausible constitutive surfaces.

### 4.11 Zener–Hollomon Correlation

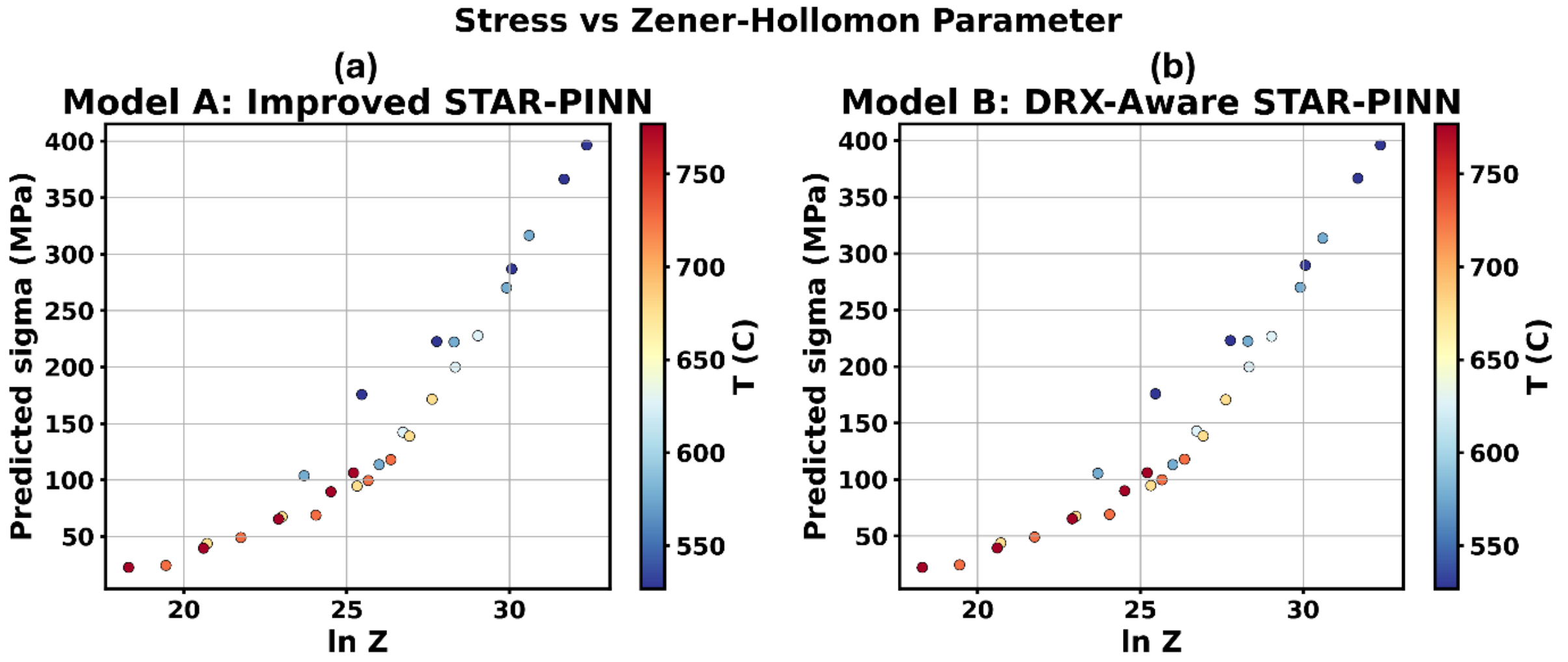

*Figure 13: The predicted flow stress at a fixed reference strain (median strain value) against the Zener–Hollomon parameter ln Z = ln($\dot{\varepsilon}$) + Q/(RT), with individual points colour-coded by temperature for (a) Improved STAR-PINN and (b) DRX-Aware STAR-PINN models.*

Figure 13 presents the relationship between the predicted flow stress and the Zener–Hollomon parameter, ln Z, for the Improved STAR-PINN and DRX-Aware STAR-PINN. Both models exhibit a clear and consistent monotonic increase in predicted flow stress with increasing ln Z, demonstrating that the learned constitutive response preserves the expected dependence of hot-deformation behavior on temperature and strain rate. At lower ln Z values (approximately 18–23), the predicted stresses remain relatively low, generally below 100 MPa, whereas increasing ln Z produces a progressive increase in flow stress, reaching approximately 400 MPa at ln Z approx 32. This behavior is consistent with the expected reduction in thermally activated recovery at lower temperatures and higher strain rates, which increases deformation resistance. The temperature color scale further confirms the expected metallurgical trend: higher temperatures correspond to lower ln Z and lower predicted stresses, while lower temperatures are associated with higher ln Z and greater flow stress. Importantly, the two models produce very similar stress ln Z distributions across the entire investigated range, with no pronounced separation between their predictions. This indicates that incorporating the DRX branch does not disrupt the fundamental Arrhenius/Zener–Hollomon dependence learned by the constitutive framework. Rather, the DRX-aware formulation provides an additional mechanism for representing recrystallization-mediated softening while retaining the same underlying thermomechanical trends.

### 4.12 Transferability to Other Alloy Compositions

A natural question is whether the proposed STAR-PINN framework would require complete retraining when applied to titanium alloys with different compositions, or whether the learned representations can be transferred. The current model is trained solely on flow-stress data from the analyzed Mo-rich α+β titanium alloy (Ti–6Al–4Mo–1V–0.1Si), meaning its learned constitutive representation is likely to show some level of composition dependence, especially when the alloy demonstrates significantly varying phase stability, β-transus temperature, or deformation mechanisms. Hence, for alloys that are compositionally distant, full or significant retraining and recalibration would probably be required. Nonetheless, the modular design of the suggested framework offers a possible pathway for transfer learning pertaining to compositionally similar alloys. Specifically, the shared residual encoder can be preserved as a preliminary representation of the nonlinear thermomechanical response, whereas the stress and DRX prediction heads can be optimized with a limited alloy-specific dataset. The adjustable JMAK parameters, k and n, along with the Arrhenius parameters, can also be re-calibrated to reflect variations in recrystallization kinetics and thermally driven deformation. This method is anticipated to be most suitable for closely related Mo-rich, near-α, or near-β titanium alloys treated within similar temperature and strain-rate ranges, where the fundamental hardening–softening and strain-rate-sensitive behavior remains qualitatively alike. For alloys featuring notably distinct compositions or primary deformation mechanisms, the physical constraints and constitutive relations would necessitate thorough re-assessment and re-calibration, and the transfer of the encoder may offer minimal advantages. Importantly, transfer learning across alloy compositions was not explicitly evaluated in the present work; therefore, we do not claim demonstrated cross-composition transferability. Rather, the modular architecture establishes a practical basis for such an approach, which should be systematically validated in future studies by freezing and fine-tuning different portions of the network using independent datasets from compositionally related titanium alloys.

## Conclusion

This research developed and assessed two Stacked Residual Physics-Informed Neural Networks (STAR-PINNs) for modeling the constitutive behavior of hot deformation in a Mo-rich α+β titanium alloy (Ti–6Al–4Mo–1V–0.1Si) over temperatures from 800 to 1050 °C and strain rates from 0.01 to 10 $s^{-1}$. The subsequent key conclusions are derived:

- The Improved STAR-PINN achieved RMSE = 11.82 MPa, MAE = 4.91 MPa, and $R^2$ = 0.9847, while the DRX-Aware STAR-PINN achieved RMSE = 11.62 MPa, MAE = 4.73 MPa, and $R^2$ = 0.9852. These findings verify that the physics-constrained, residual-stacked architecture generalizes successfully throughout a wide thermomechanical processing range.

- Physics-Informed constraints applied through automatic differentiation, such as thermal softening ($\frac{\partial \sigma}{\partial \mathrm{T}} < 0$), strain-rate sensitivity ($\frac{\partial \sigma}{\partial \ln(\dot{\varepsilon})} > 0$), pre-peak strain hardening, post-peak flow softening, and curvature regularization, directed both models toward thermodynamically consistent constitutive behavior without needing a direct analytical constitutive equation.
- The DRX-Aware STAR-PINN, utilizing its dual-output structure and latent-feature integration, incorporated the mechanistic relationship between recrystallization-driven softening and overall flow resistance directly into the network design. The JMAK–Avrami regularization, DRX monotonicity constraint, DRX–softening coupling loss, saturation constraint, and Arrhenius-consistency penalty together delivered physically consistent DRX fraction forecasts that showcased sigmoidal kinetics, accurate strain-rate sequencing, and saturation behavior approaching $X_{DRX} \rightarrow 1$.
- The JMAK–Avrami parameters $k$and $n$were treated as learnable, bounded parameters rather than being fixed a priori. In the main training run, they converged to $k = 2.03$ and $n = 1.99$, while five-fold cross-validation yielded $k = 2.01 \pm 0.01$ and $n = 1.99 \pm 0.002$. Their close agreement with the conventional initial values of $k = n = 2.0$demonstrates that these parameters are appropriate for the present alloy while allowing their values to be inferred directly from the available deformation data.
- Monte Carlo Dropout uncertainty quantification indicated that predictive confidence is greatest during the steady-state deformation phase and least close to the peak-stress transition under high strain rates and low temperatures, a physically relevant uncertainty distribution that is immediately usable in finite-element process simulation.
- Gradient-based feature importance analysis showed that the DRX-Aware architecture attained a 25.7 % decrease in raw strain sensitivity compared to the Improved STAR-PINN, reallocating predictive dependence towards temperature and strain rate. This verifies that explicit DRX conditioning partially separates the constitutive surface decomposition in a way that aligns with the Zener–Hollomon framework.
- Zener–Hollomon correlation analysis confirmed that both models demonstrate the anticipated monotonic rise of flow stress with ln Z, with the DRX-Aware STAR-PINN yielding a distinctly smoother and more reliable path through the intermediate ln Z range (≈26–29) where hardening and DRX softening compete.

- Sensitivity analysis of the post-peak softening weight $W_{\text{soft}}$and DRX–softening coupling weight $W_{\text{couple}}$, varied independently over 0.5×, 1×, and 2× their reference values, produced test-set RMSE values between 11.82 and 12.28 MPa and $R^2$ values between 0.9834 and 0.9847. The less-than-4% variation in RMSE demonstrates that the model is relatively robust to moderate changes in these physics-loss weights and is not dependent on an exceptionally narrow weighting configuration. The best-performing combination was $W_{\text{soft}} = 1\,\times$and $W_{\text{couple}} = 2\,\times$, with RMSE = 11.82 MPa.

Future work should extend the DRX-Aware framework to include grain size prediction as a third output head, incorporate EBSD microstructural measurements as additional soft training targets, and validate the framework on other near-β titanium alloys to assess generalizability across alloy compositions.

**Acknowledgments**

The authors are grateful to the Director of VNIT Nagpur for providing the necessary facilities and offering constant encouragement for this research.

**Conflict of Interest Statement**

The authors declare that they have no known competing financial interests or personal relationships that could have influenced the work reported in this paper.

**Credit Authorship**

Prashil S. Joshi: Conceptualization, Methodology, Visualization, Investigation, Writing – original draft. Diksha Mahadule: Methodology, Visualization, Writing – review & editing. Rajesh K. Khatirkar: Supervision, Writing – review & editing, Resources.

**Data Availability Statement**

The data supporting the findings of this study are available from the corresponding author upon reasonable request.